# CP VIOLATION IN THE K AND B SYSTEMS[†]

Boris Kayser

Fermi National Accelerator Laboratory
Batavia, IL 60510 USA
and
National Science Foundation[*]
4201 Wilson Boulevard
Arlington, VA 22230 USA

## Abstract

Although CP violation was discovered more than thirty years ago, its origin is still unknown. In these lectures, we describe the CP-violating effects which have been seen in K decays, and explain how CP violation can be caused by the Standard Model weak interaction. The hypothesis that this interaction is indeed the origin of CP violation will be incisively tested by future experiments on B and K decays. We explain what quantities these experiments will try to determine, and how they will be able to determine them in a theoretically clean way. To clarify the physics of the K system, we give a phase-convention-free description of CP violation in this system. We conclude by briefly exploring whether electric dipole moments actually violate CP even if CPT invariance is not assumed.

---

[†] To appear in the Proceedings of the Summer School in High Energy Physics and Cosmology, held at the International Center for Theoretical Physics, Trieste, June-July 1995.

[*] Permanent Address

## 1. Preamble

Being lovers of symmetry, we are tempted to expect that the behavior of a physical system will not change at all if we replace every particle in it by its antiparticle. However, in some processes nature violates this expected matter-antimatter symmetry. Particularly interesting is the violation of invariance under CP, the combined action of charge conjugation C and parity P. Before we turn to CP, let us first briefly consider the simplest operation which replaces a particle by its antiparticle, namely C.

The effect of C on a particle $f(\vec{p},\lambda)$ of momentum $\vec{p}$ and helicity $\lambda$ is given by

$$C|f(\vec{p},\lambda)\rangle = \eta_C|\bar{f}(\vec{p},\lambda)\rangle \ , \tag{1.1}$$

where $\bar{f}$ is the antiparticle of f, and $\eta_C$ is a phase factor. Note that C does not alter a particle's momentum or helicity.

It has long been known that some processes are not invariant under C. Consider, for example, the decay $\pi \to \mu\nu$. Invariance under C would require that the muons produced in $\pi^+ \to \mu^+ + \nu$ and $\pi^- \to \mu^- + \bar{\nu}$ have identical helicity. But it is found that actually they have opposite helicity: the $\mu^+$ in $\pi^+$ decay is always left-handed, while the $\mu^-$ in $\pi^-$ decay is always right-handed. Thus, $\pi \to \mu\nu$ is not invariant under C.

A somewhat more subtle operation which replaces a particle by its antiparticle is CP. The effect of CP on $f(\vec{p},\lambda)$ is given by

$$CP|f(\vec{p},\lambda)\rangle = \eta_{CP}|\bar{f}(-\vec{p},-\lambda)\rangle \ , \tag{1.2}$$

Here, the momentum and helicity reversals are due to the action of P, and $\eta_{CP}$ is a phase factor.

From Eq. (1.2), the CP-mirror image of the decay $\pi^+ \to \mu^+_{LH} + \nu$, where the subscript LH reminds us that the $\mu^+$ has left-handed helicity, is the decay $\pi^- \to \mu^-_{RH} + \bar{\nu}$, in which the $\mu^-$ has right-handed helicity. These two decays are the ones actually observed, and they have equal rates. Thus, $\pi \to \mu\nu$, while not invariant under C, *is* invariant under CP.

One might wonder whether perhaps *all* processes, even those which are not invariant under C, are nevertheless invariant under CP. The decays of the neutral K mesons have taught us that this is not the case. Let us turn, then, to the phenomenology of the neutral kaon system.

## 2. CP Violation in the Neutral K System

Let us consider neutral kaons at rest. We choose our phase conventions in this Section so that

$$CP|K^0\rangle = +|\overline{K^0}\rangle \ . \tag{2.1}$$

Simple field theory then implies that

$$CP|\overline{K^0}\rangle = +|K^0\rangle \ . \tag{2.2}$$

In Section 5, we shall free ourselves of this phase convention, and adopt a convention-free formalism.

While $K^0$ and $\overline{K^0}$ have opposite strangeness, the weak interactions do not conserve strangeness, and so they mix $K^0$ and $\overline{K^0}$. The time evolution of a neutral kaon is then described by a two-component Schrödinger equation of the form

$$i\frac{\partial}{\partial t}\begin{bmatrix} a(t) \\ \bar{a}(t) \end{bmatrix} = \mathcal{M}\begin{bmatrix} a(t) \\ \bar{a}(t) \end{bmatrix} \ . \tag{2.3}$$

Here, a(t) is the amplitude for us to have a $K^0$ at time t, and $\bar{a}(t)$ is the amplitude for us to have a $\overline{K^0}$. The quantity M is a 2x2 matrix,

$$\mathcal{M} = \begin{bmatrix} \langle K^0|\mathcal{M}|K^0\rangle & \langle K^0|\mathcal{M}|\overline{K^0}\rangle \\ \langle \overline{K^0}|\mathcal{M}|K^0\rangle & \langle \overline{K^0}|\mathcal{M}|\overline{K^0}\rangle \end{bmatrix} \equiv \begin{bmatrix} \mathcal{M}_{11} & \mathcal{M}_{12} \\ \mathcal{M}_{21} & \mathcal{M}_{22} \end{bmatrix} \ , \tag{2.4}$$

known as the neutral K mass matrix, which serves as the effective Hamiltonian for a neutral kaon at rest. Since a kaon disappears with time through its decay, $\mathcal{M}$ is non-Hermitean.

We shall assume that the world is CPT invariant. It is readily shown that this invariance implies that

$$\mathcal{M}_{11} \equiv \langle K^0|\mathcal{M}|K^0\rangle = \langle\overline{K^0}|\mathcal{M}|\overline{K^0}\rangle \equiv \mathcal{M}_{22} \ . \tag{2.5}$$

Writing $\mathcal{M}_{11} \equiv M - i\Gamma/2$, and $\mathcal{M}_{22} \equiv \overline{M} - i\overline{\Gamma}/2$, we see that the real part of Eq. (2.5), $M = \overline{M}$, is an example of the well-known CPT requirement that a particle and its antiparticle have the same mass. The imaginary part, $\Gamma = \overline{\Gamma}$, is an example of the requirement that they have the same total width or, equivalently, the same lifetime.

Assume for the moment that CP invariance holds. Then our effective Hamiltonian obeys $(CP)^\dagger \mathcal{M}(CP) = \mathcal{M}$, and we have

$$\mathcal{M}_{12} \equiv \langle K^0 | \mathcal{M} | \overline{K^0} \rangle = \langle K^0 | (CP)^\dagger \mathcal{M}(CP) | \overline{K^0} \rangle$$

$$= \langle (CP)K^0 | \mathcal{M} | (CP)\overline{K^0} \rangle = \langle \overline{K^0} | \mathcal{M} | K^0 \rangle \equiv \mathcal{M}_{21} \ . \qquad (2.6)$$

Thus, from Eqs. (2.5) and (2.6), $\mathcal{M}$ has the form

$$\mathcal{M} = \begin{bmatrix} X & Y \\ Y & X \end{bmatrix} \ . \qquad (2.7)$$

The eigenstates of $\mathcal{M}$—the mass eigenstates of the neutral K system—are then $\frac{1}{\sqrt{2}}\begin{bmatrix} 1 \\ 1 \end{bmatrix}$ and $\frac{1}{\sqrt{2}}\begin{bmatrix} 1 \\ -1 \end{bmatrix}$. That is, they are the CP eigenstates

$$|K_1\rangle = \frac{1}{\sqrt{2}}\left[|K^0\rangle + |\overline{K^0}\rangle\right] \qquad (2.8)$$

and

$$|K_2\rangle = \frac{1}{\sqrt{2}}\left[|K^0\rangle - |\overline{K^0}\rangle\right] \ . \qquad (2.9)$$

Note that $|K_1\rangle$ and $|K_2\rangle$ have opposite CP parity:

$$CP|K_{1(2)}\rangle = {}_{(-)}^{+}|K_{1(2)}\rangle \ . \qquad (2.10)$$

Now, experimentally, the two mass eigenstates of the neutral K system are the K-short $K_S$, with a short lifetime $\tau_S = (0.8926 \pm 0.0012) \times 10^{-10}$ sec, and the K-long $K_L$, with a much longer lifetime $\tau_L = (5.17 \pm 0.04) \times 10^{-8}$ sec. Essentially all $K_S$ decays are to $\pi^+\pi^-$ or $\pi^0\pi^0$. It is easy to show that both of these final states have CP = +1. Thus, if CP invariance and the associated CP conservation law hold, $K_S$ must be $K_1$. Then $K_L$ must be $K_2$. But then, since $K_2$ has CP = –1, the decays $K_L \to \pi^+\pi^-$ and $K_L \to \pi^0\pi^0$ are forbidden. Nevertheless, these decays do occur.[1] Thus, CP is violated in neutral K decays.

To be sure, the observed violation is small. The amplitudes for the CP-violating decay $K_L \to \pi^+\pi^-$ and the CP-conserving one $K_S \to \pi^+\pi^-$ are in the ratio[2]

$$\frac{\left|\langle \pi^+\pi^-|T|K_L\rangle\right|}{\left|\langle \pi^+\pi^-|T|K_S\rangle\right|} = (2.285 \pm 0.019) \times 10^{-3} \ . \qquad (2.11)$$

Similarly,[2]

$$\frac{\left|\left\langle \pi^0\pi^0 \middle| T \middle| K_L \right\rangle\right|}{\left|\left\langle \pi^0\pi^0 \middle| T \middle| K_S \right\rangle\right|} = (2.275 \pm 0.019) \times 10^{-3} \quad . \tag{2.12}$$

Nevertheless, this violation of CP is nonvanishing. Like C, CP is a symmetry which is not always respected.

In addition to $K_L \to \pi\pi$, other CP-violating effects have been seen in the neutral kaon system. One of these is found in the semileptonic decays $K_L \to \pi l \nu$, where $l$ is an e or a $\mu$. When CP invariance holds, the $K_L$ is a CP eigenstate. Then the CP-mirror image of the decay $K_L \to \pi^- l^+ \nu$ is $K_L \to \pi^+ l^- \bar{\nu}$. To be sure, CP reverses the momenta and helicities of all the outgoing particles, but that is irrelevant when we integrate over these variables to get the full rate for decay into the particles under consideration. Thus, when CP invariance holds, we require that $\Gamma(K_L \to \pi^- l^+ \nu) = \Gamma(K_L \to \pi^+ l^- \bar{\nu})$. However, it is found experimentally that[2]

$$\delta \equiv \frac{\Gamma\left(K_L \to \pi^- l^+ \nu\right) - \Gamma\left(K_L \to \pi^+ l^- \bar{\nu}\right)}{\text{"} \qquad + \qquad \text{"}}$$

$$= 3.27 \pm 0.12 \times 10^{-3} \quad . \tag{2.13}$$

Further observed CP-violating effects will be discussed in Section 5.

## 3. The Origin of CP Violation in the Standard Model

All CP-violating effects observed to date have been seen in the decays of neutral kaons. These decays are known to be due to the weak interaction. Therefore, it is natural to speculate that CP violation may well be an effect of the weak interaction. This is the possibility that we shall emphasize here.

The weak interaction is very successfully described by the so-called Standard Model (SM). In the SM, the weak interaction is carried by the charged weak boson W, and the neutral weak boson Z. These bosons couple to the leptons and to the three generations, or families, of quarks:

| Generation: | 1 | 2 | 3 |
|---|---|---|---|
| Charge = 2/3<br>Charge = –1/3 | $\begin{pmatrix} u \\ d \end{pmatrix}$ | $\begin{pmatrix} c \\ s \end{pmatrix}$ | $\begin{pmatrix} t \\ b \end{pmatrix}$ |

$\longrightarrow$ Mass

As indicated, the quarks in the third generation are the heaviest ones, those in the second generation are lighter, and those in the first generation are lighter still. In a typical Feynman diagram for a hadronic weak decay, a relatively heavy quark decays to lighter ones via W exchange. This is illustrated in Fig. 1, in which a $K_S$ decays via its $|\overline{K^0}\rangle$ ( $= |s\bar{d}\rangle$) component into $\pi^+\pi^-$. The diagram of Fig. 1 entails the W-mediated quark decay $s \to u\bar{u}d$.

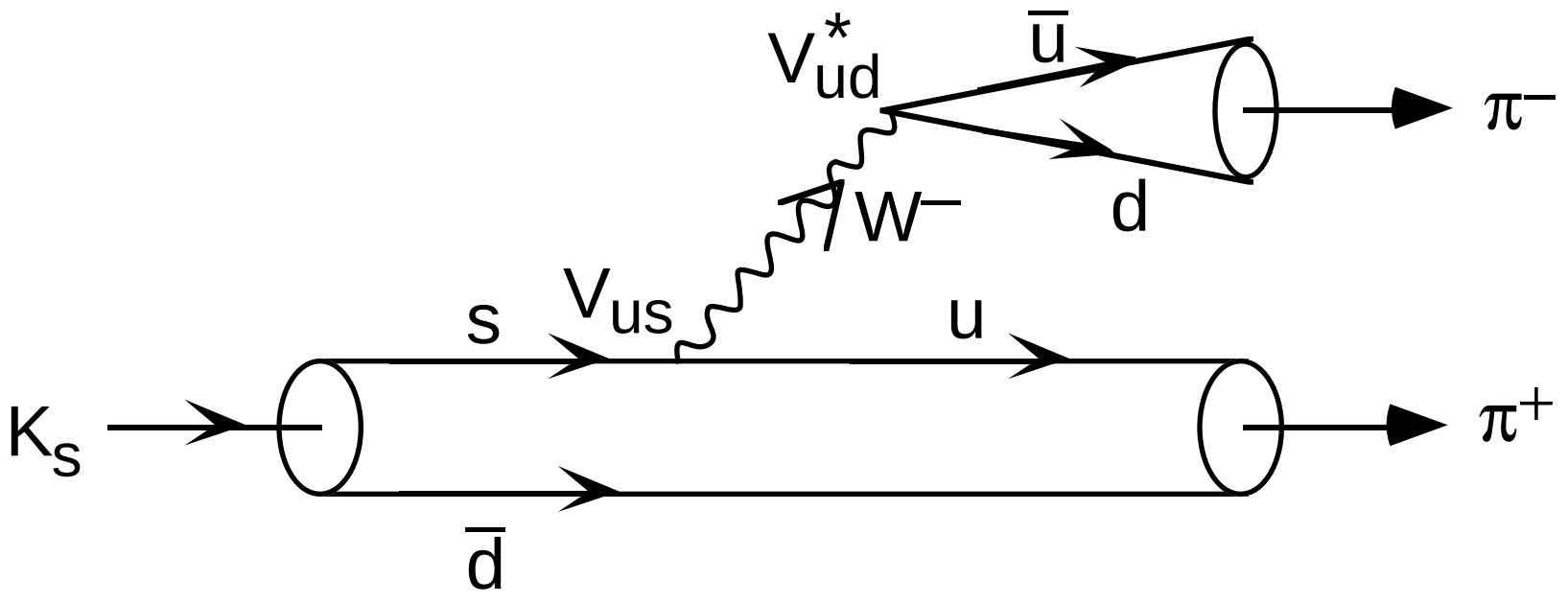

Figure 1. One of the diagrams for $K_S \to \pi^+\pi^-$.

According to the SM, the coupling of the W boson to the quarks is given by the Hamiltonian

$$\mathcal{H}_W = \frac{g}{\sqrt{2}} W^\mu \sum_{\substack{\alpha=u,c,t \\ i=d,s,b}} V_{\alpha i} \overline{\alpha_L} \gamma_\mu i_L + \frac{g}{\sqrt{2}} W^{\mu^\dagger} \sum_{\substack{\alpha \\ i}} V^*_{\alpha i} \overline{i_L} \gamma_\mu \alpha_L \quad . \tag{2.14}$$

Here, g is a real overall coupling strength, W is the W boson field, $\alpha_L \equiv \frac{1-\gamma_5}{2}\alpha$ is the left-handed projection of the quark field $\alpha$ and similarly for $i_L$, and the numerical coefficients $V_{\alpha i}$ are the elements of the Cabibbo-Kobayashi-Maskawa (CKM) quark mixing matrix

$$V = \begin{pmatrix} V_{ud} & V_{us} & V_{ub} \\ V_{cd} & V_{cs} & V_{cb} \\ V_{td} & V_{ts} & V_{tb} \end{pmatrix} \quad . \tag{2.15}$$

Note from Eq. (2.14) that any of the three negatively-charged quarks i can turn into any of the three positively-charged ones $\alpha$ by emitting a $W^-$, the amplitude for its doing so being proportional to $V_{\alpha i}$. Thus, the off-diagonal elements of V describe transitions in which a quark in one family turns into a quark in another family. Hence, these elements mix the families, earning for V the name "quark mixing matrix".

The SM coupling of the Z boson to the quarks is described by the Hamiltonian

$$\mathcal{H}_Z = \frac{g}{\cos\theta_W} Z^\mu \sum_{\substack{q=u,c,t,\\ d,s,b}} \left\{ \left[ I_3(q_L) - Q(q)\sin^2\theta_W \right] \bar{q}_L \gamma_\mu q_L - Q(q)\sin^2\theta_W \bar{q}_R \gamma_\mu q_R \right\} \quad . \tag{2.16}$$

Here, $\theta_W$ is the Weinberg angle, $I_3(q_L)$ is the weak isospin of the left-handed quark $q_L$, $Q(q)$ is the electric charge of q, and $q_R \equiv \frac{(1+\gamma_5)}{2} q$ is the right-handed projection of the quark field q.

Under CP, the term $\frac{g}{\sqrt{2}} W^\mu V_{\alpha i} \bar{\alpha}_L \gamma_\mu i_L$ in the W-quark interaction $\mathcal{H}_W$ transforms as

$$(CP)\frac{g}{\sqrt{2}} W^\mu V_{\alpha i} \bar{\alpha}_L \gamma_\mu i_L (CP)^{-1} = \left[ \eta^*(W)\eta(\alpha)\eta^*(i) \right] \frac{g}{\sqrt{2}} W^{\mu\dagger} V_{\alpha i} \bar{i}_L \gamma_\mu \alpha_L \quad . \tag{2.17}$$

Here, $\eta(W)$, etc. are phases, and we may choose $[\eta^*(W)\eta(\alpha)\eta^*(i)] = 1$. Then

$$\begin{aligned} \mathcal{H}_W &= \frac{g}{\sqrt{2}} W^\mu \sum_{\alpha,i} V_{\alpha i} \bar{\alpha}_L \gamma_\mu i_L + \frac{g}{\sqrt{2}} W^{\mu\dagger} \sum_{\alpha,i} V^*_{\alpha i} \bar{i}_L \gamma_\mu \alpha_L \\ &\xrightarrow[CP]{} \frac{g}{\sqrt{2}} W^{\mu\dagger} \sum_{\alpha,i} V_{\alpha i} \bar{i}_L \gamma_\mu \alpha_L + \frac{g}{\sqrt{2}} W^\mu \sum_{\alpha,i} V^*_{\alpha i} \bar{\alpha}_L \gamma_\mu i_L \end{aligned} \quad . \tag{2.18}$$

We see that $\mathcal{H}_W$ is CP-invariant if, and only if, V is real, or can be made real by changing the phase conventions for the quark fields.

The analogue of Eq. (2.17) for the terms in the Z-quark interaction $\mathcal{H}_Z$ states that each of these terms transforms back into itself under CP. Thus, $\mathcal{H}_Z$ is necessarily CP-invariant.

We conclude that in the SM weak interaction, CP violation can arise only if some of the numbers $V_{\alpha i}$ are complex. How their complexity can produce physical CP-violating effects will be explained shortly.

### 3.1. The CKM Matrix

The SM requires that the CKM quark mixing matrix be unitary. Apart from this unitarity, the matrix is not predicted, so its elements must be determined experimentally.

How many independent parameters are needed to determine fully the CKM matrix V? In answering this question, we must bear in mind that some of the complex phases which V may contain are not physically meaningful. To see this, note from Eq. (2.14) that, apart from irrelevant factors, $V_{\alpha i}$ is just the amplitude $\langle\alpha|\mathcal{H}_W|i\rangle$ for the quark transition $i \to \alpha$ through W emission. Thus, if the arbitrary relative phase of the i and $\alpha$ quarks is changed, the phase of $V_{\alpha i}$ will change correspondingly. Redefining the down-type quark i by $|i\rangle \to e^{i\theta}|i\rangle$ multiplies the i column of V by $e^{i\theta}$. Similarly, phase redefining the up-type quark $\alpha$ multiplies the $\alpha$ row of V by a phase factor. Hence, without changing the physics, we may multiply any column or row of V by a phase factor, or carry out any number of such operations. We may use these operations to remove from V five phases corresponding to the five relative phases of the six quarks, leaving five of the elements of V real. We may do this, for example, by multiplying each of the columns of V by a phase factor chosen to make its bottom element real, and then multiplying each of the top two rows by a phase factor chosen to make its rightmost element real.

Mindful of this freedom to remove at least some phases from V, let us now suppose that there are, not just three doublet quark families, but N of them, so that V becomes an NxN matrix. How many parameters are necessary to determine it completely? Before constraints are imposed, $2N^2$ real numbers are required to fully specify the $N^2$ complex elements of V. But the unitarity of V demands that the sum of the absolute squares of the elements in any of its columns be unity—a demand that imposes N constraints. Furthermore, unitarity demands that any two of the columns of V be orthogonal. Now, there are $N(N-1)/2$ pairs of columns, and the equation expressing the orthogonality of any pair has both a real and an imaginary part. Thus, orthogonality of columns imposes $N(N-1)$ constraints. Hence, the most general NxN unitary matrix depends on $2N^2 - N - N(N-1) = N^2$ real parameters. Now, in the N quark families there are 2N quarks, with $2N-1$ relative phases. Thus, $2N-1$ phases in V are not physically meaningful, and may be removed by phase redefinitions of the quarks or, equivalently, by multiplying columns and rows of V by phase factors. Hence, the number of physically meaningful independent real parameters in V is $N^2-(2N-1) = (N-1)^2$.

One possible choice for these parameters is mixing angles (parameters which would be present even if V were real) and phases. To calculate the number of mixing angles on which V depends, imagine that it is real. It is then an orthogonal (i.e., a rotation) matrix. It contains $N^2$ real elements, subject to N constraints expressing the requirement that each of its columns be a vector of unit length, and $N(N-1)/2$ constraints expressing the requirement that any pair of its columns be orthogonal. Thus, V depends on $N^2 - N - N(N-1)/2 = N(N-1)/2$ mixing angles.

In summary, the complex NxN quark mixing matrix depends on $(N-1)^2$ parameters. If we take these to be mixing angles and phases, $N(N-1)/2$ of them are mixing angles, so that $(N-1)^2 - N(N-1)/2 = (N-1)(N-2)/2$ of them are phases. Note from this result that there are *no* physically significant phases in the mixing matrix unless $N \geq 3$.[3] Had there been fewer than three quark families, it would have been impossible for the weak interaction, as described in the SM, to violate CP.

(It is instructive and easy to explicitly construct the most general unitary quark mixing matrix for the case N = 2, and show that all phases can be removed from this matrix by multiplying its rows and columns by phase factors. Since this leaves the matrix real, it cannot violate CP.)

Although there are in reality (at least) three quark families, the fact that the quark mixing matrix cannot violate CP in a world with only two families has an important consequence. Namely, it implies that CP violation in K decays will be small, as observed, even if the complex phases in the true 3x3 mixing matrix are large. Furthermore, it tells us where we must look if we wish to see large CP-violating effects.[4] To see why it implies that CP violation in K decays will be small, note that, as illustrated in Fig. 1, the dominant diagrams for K decays involve only quarks from the first two families. The quarks t and b are not involved. Thus, in first approximation, K decays "do not know" that there are not just two, but three quark generations. In this approximation, K decay processes do not contain enough physics to violate CP. Now, when one goes beyond the first approximation, one finds that K decays do involve the quarks of the third generation in several ways, so that these reactions can (and do!) violate CP. However, because one must go beyond the leading approximation before the third generation quarks come into the picture, CP violation in K decays is small.

From this discussion, it is clear that if we wish to see large CP-violating effects coming from the CKM matrix, we must look for them in processes which involve, even in leading approximation, quarks from all three generations. To this end, new facilities are being built and new experiments are being developed which will study CP violation in the decays of the B or beauty mesons. The B mesons and their quark content are

$$B^+ = [\bar{b}u] \qquad B^- = [b\bar{u}]$$
$$B_d = [\bar{b}d] \qquad \overline{B_d} = [b\bar{d}]$$
$$B_s = [\bar{b}s] \qquad \overline{B_s} = [b\bar{s}]$$
$$B_c = [\bar{b}c] \qquad \overline{B_c} = [b\bar{c}] \ .$$

In a typical B decay, the heavy b or $\bar{b}$ quark in the B—a quark of the third generation—decays down to lighter quarks in the first and/or second generations. Often, all three generations are involved. Thus, CP violation can be large.

As we shall see shortly, the CP-violating effects in B decays can also yield *clean* information on the phases in the CKM matrix. Thus, the study of these effects will be a very good test of whether these phases are indeed the origin of CP violation.

## 4. CP Violation in the B System

The effects to be sought in the B system are CP-violating inequalities between the rates for CP-mirror-image decays. When CP invariance holds, the amplitude $\langle f|T|i\rangle$ for the decay of any initial state i into any final one f obeys

$$\langle f|T|i\rangle = \langle CP[f]|T|CP[i]\rangle \ . \tag{4.1}$$

Thus, for example, any inequality between the rates for the CP-mirror-image decays $B^+ \to f$ and $B^- \to \bar{f}$, where $\bar{f} \equiv CP[f]$ is the CP-mirror image of the final state f, is a violation of CP invariance. It is violations of this general sort, which are B-system analogues of the K-system asymmetry δ of Eq. (2.13), which will be sought.

As we noted earlier, CPT invariance, which we assume to hold exactly, requires that any unstable particle and its antiparticle have the same total width. Thus, if there is some final state f for which, in violation of CP, $\Gamma[B^+ \to f] > \Gamma[B^- \to \bar{f}]$, then there must be some other final state (or states) f′ for which $\Gamma[B^+ \to f'] < \Gamma[B^- \to \bar{f}']$. Otherwise, the CPT constraint that $\Gamma_{total}[B^+] = \Gamma_{total}[B^-]$ could not be satisfied.

The complex phases in the CKM matrix, like complex phases anywhere in quantum mechanics, lead to physical consequences only through interferences between amplitudes. In particular, it is through interferences that the CKM phases produce CP violation. How they do this is nicely illustrated by the comparison between the CP-mirror-image processes $B^+ \to f$ and $B^- \to \bar{f}$. Suppose that the weak decay $B^+ \to f$ receives contributions from two Feynman diagrams. Each of these diagrams is proportional, like the diagram of Fig. 1, to some product of CKM elements. Thus, the amplitude a for the first diagram has the form

$$a = Me^{i\delta^f_{CKM}}e^{i\alpha_S} \ , \tag{4.2}$$

where M is the magnitude of a, $\delta^f_{CKM}$ is the phase of the product of CKM elements to which the diagram is proportional, and $\alpha_S$ is a phase arising from strong interaction effects such as final-state rescattering. Similarly, the amplitude a′ for the second diagram has the form

$$a' = M' e^{i\delta'^f_{CKM}} e^{i\alpha'_S} \quad , \tag{4.3}$$

where M' ≡ |a′|, $\delta'^f_{CKM}$ is the phase of the product of CKM elements to which the second diagram is proportional, and $\alpha'_S$ is the strong-interaction phase of this diagram. The rate for $B^+ \to f$ is then

$$\begin{aligned} \Gamma\left[B^+ \to f\right] &= \left| M e^{i\delta^f_{CKM}} e^{i\alpha_S} + M' e^{i\delta'^f_{CKM}} e^{i\alpha'_S} \right|^2 \\ &= M^2 + M'^2 + 2MM'\cos(\varphi + \varphi_S) \end{aligned} \quad , \tag{4.4}$$

where $\varphi \equiv \delta^f_{CKM} - \delta'^f_{CKM}$ is the relative CKM phase of the two amplitudes, and $\varphi_S \equiv \alpha_S - \alpha'_S$ is their relative strong-interaction phase.

Now, the diagrams for the CP-mirror-image decay $B^- \to \bar{f}$ are, of course, the same as those for $B^+ \to f$, except that every quark (antiquark) is replaced by its antiquark (quark). From $\mathcal{H}_W$, Eq. (2.14), we see that, owing to this replacement, every CKM element appearing in a diagram for $B^+ \to f$ is replaced by its complex conjugate in the corresponding diagram for $B^- \to \bar{f}$. However, apart from CKM phases, the SM weak interaction of Eqs. (2.14) and (2.16) is completely CP invariant, as is the SM strong interaction. Thus, apart from the reversal of its CKM phase, the amplitude of a diagram does not change at all when we go from $B^+ \to f$ to $B^- \to \bar{f}$. Hence, the rate for $B^- \to \bar{f}$ is

$$\begin{aligned} \Gamma\left[B^- \to \bar{f}\right] &= \left| M e^{-i\delta^f_{CKM}} e^{i\alpha_S} + M' e^{-i\delta'^f_{CKM}} e^{i\alpha'_S} \right|^2 \\ &= M^2 + M'^2 + 2MM'\cos(-\varphi + \varphi_S) \end{aligned} \quad . \tag{4.5}$$

Comparing Eqs. (4.4) and (4.5), we see that when CKM phases are present, the two interfering amplitudes can have a different relative phase in $B^- \to \bar{f}$ than they do in $B^+ \to f$. As a result, $\Gamma[B^- \to \bar{f}]$ and $\Gamma[B^+ \to f]$ can differ, in violation of CP.

To test the SM of CP violation, one would like not only to observe a CP-violating inequality between $\Gamma[B^+ \to f]$ and $\Gamma[B^- \to \bar{f}]$, but also to determine the CKM phase φ. Of course, $\Gamma[B^+ \to f]$ and $\Gamma[B^- \to \bar{f}]$ are only *two* measurable quantities, and as we see from Eqs. (4.4) and (4.5), they depend on *four* parameters: M, M′, φ, and $\varphi_S$. Thus, by themselves, they cannot determine $\varphi$. Consequently, in general, a clean test of the SM of CP violation is not possible in decays of charged B mesons. (To be sure, in the exceptional cases where M and M′ can be determined independently of $\Gamma[B^+ \to f]$ and $\Gamma[B^- \to \bar{f}]$, the measurement of these two decay rates determines $\sin^2\varphi$, up to a two-fold ambiguity, and so does provide a test of the SM.[5])

### 4.1. Decays of Neutral B Mesons

In decays of *neutral* B mesons, a clean test of the SM of CP violation *is* possible. To see why, let us discuss the $B_d - \overline{B_d}$ system; the $B_s - \overline{B_s}$ system behaves similarly.

The key feature of the $B_d - \overline{B_d}$ system is the fact that the $B_d$ and $\overline{B_d}$ mix. In the SM, they do so largely as a result of the box diagram in Fig. 2. The phase of the mixing amplitude $A(B_d \to \overline{B_d})$ is then

$$\arg\left[A\left(B_d \to \overline{B_d}\right)\right] = \arg\left[\left(V_{td}V_{tb}^*\right)^2\right] \equiv -2\delta_{CKM}^m \quad . \tag{4.6}$$

We shall refer to $\delta_{CKM}^m$ (where m stands for mixing) as the $B_d - \overline{B_d}$ mixing phase.

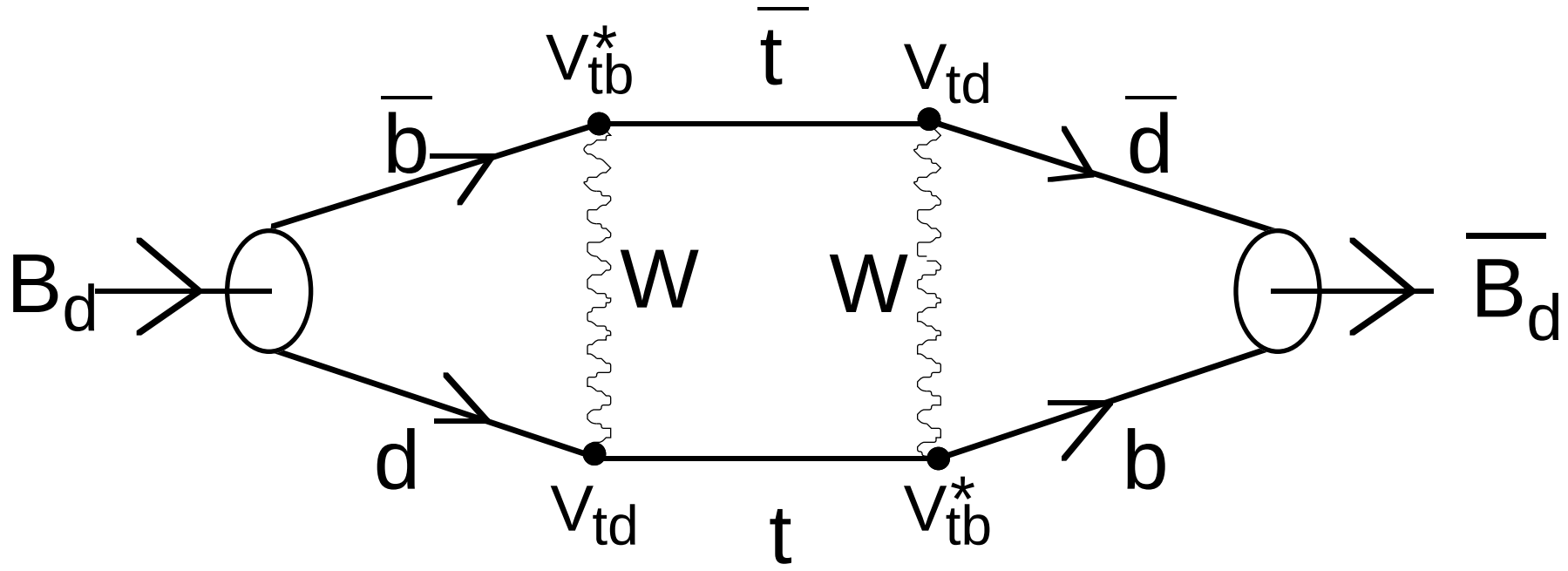

Figure 2. The SM box diagram for $B_d - \overline{B_d}$ mixing.

Time evolution in the $B_d - \overline{B_d}$ system is described by a two-component Schrödinger equation, just like the one for neutral kaons, Eq. (2.3). The $B_d - \overline{B_d}$ Schrödinger equation involves a mass matrix,

$$\mathcal{M} = \begin{bmatrix} X & \mathcal{M}_{12} \\ \mathcal{M}_{21} & X \end{bmatrix} , \tag{4.7}$$

which is the $B_d - \overline{B_d}$ counterpart to the K mass matrix of Eq. (2.4). In the $B_d - \overline{B_d}$ mass matrix, the diagonal elements X are equal because of CPT, and the off-diagonal element $\mathcal{M}_{21}$ comes from the box diagram of Fig. 2. The remaining off-diago-nal element, $\mathcal{M}_{12}$, comes from a similar box diagram in which every quark (antiquark) has been replaced by its antiquark (quark). As in the case of charged B decays, this means that every CKM element has been replaced by its complex conjugate, but there have been no other changes. Since the box diagram of Fig. 2 has no strong phase (owing to the fact that the B meson is far below $\bar{t}t$ threshold), we see that

$$\mathcal{M}_{12} = \mathcal{M}_{21}(V \to V^*) = \mathcal{M}_{21}^* . \tag{4.8}$$

Let us call the mass eigenstates of the $B_d - \overline{B}_d$ system $B_{Heavy}$ ($B_H$) and $B_{Light}$ ($B_L$). From Eq. 4.7, the complex masses of these mass eigenstates—the eigenvalues of $\mathcal{M}$— are

$$\lambda_{H(L)} = X \underset{(-)}{+} \sqrt{\mathcal{M}_{12}\mathcal{M}_{21}} \equiv m_{H(L)} - \frac{i}{2}\Gamma_{H(L)} . \tag{4.9}$$

Here, $m_{H(L)}$ are the masses of $B_{H(L)}$, respectively, and $\Gamma_{H(L)}$ are their widths. Note that since $\mathcal{M}_{12}\mathcal{M}_{21}$ is real and positive, so that $\lambda_H$ and $\lambda_L$ have the same imaginary part, the widths of $B_H$ and $B_L$ are equal:

$$\Gamma_H = \Gamma_L \equiv \Gamma . \tag{4.10}$$

(To a very good approximation, this equality holds even if the SM diagram of Fig. 2 is not a good approximation to $A(B_d \to \overline{B}_d)$. This is simply because, unlike $K_S$ and $K_L$, neither B mass eigenstate has a special decay mode which is an appreciable fraction of its decays and which is unavailable to the other mass eigenstate. Thus, $B_H$ and $B_L$ have approximately equal widths.)

From Eqs. (4.7) and (4.9), the mass eigenstates $|B_{H(L)}\rangle$ are given by

$$|B_{H(L)}\rangle = \frac{1}{\sqrt{2}}\left[|B_d\rangle \underset{(-)}{+} e^{-2i\delta^m_{CKM}}|\overline{B_d}\rangle\right] . \tag{4.11}$$

Here and hereafter we assume that $\mathcal{M}_{21}$ does come from the SM diagram of Fig. 2.

Owing to the $B_d$ —$\overline{B}_d$ mixing, a neutral B at rest which at time t = 0 is a pure $|B_d\rangle$ will not remain that way. Rather, in time t it will evolve into a state $|B_d(t)\rangle$ which is a coherent superposition of $|B_d\rangle$ and $|\overline{B}_d\rangle$. From Eqs. (4.9), (4.11), and Schrödinger's equation, it is straightforward to show that

$$|B_d(t)\rangle = e^{-i\left(m - i\frac{\Gamma}{2}\right)t}\left\{c|B_d\rangle - ie^{-2i\delta^m_{CKM}} s|\overline{B_d}\rangle\right\} . \tag{4.12}$$

Here,

$$m \equiv \frac{m_H + m_L}{2} \tag{4.13}$$

is the average $B_H$, $B_L$ mass,

$$\Delta m \equiv m_H - m_L \tag{4.14}$$

is the $B_H - B_L$ mass difference, and

$$c \equiv \cos\left(\frac{\Delta m}{2}t\right), \quad s \equiv \sin\left(\frac{\Delta m}{2}t\right) \ . \tag{4.15}$$

Note from Eq. (4.12) that, before it decays into some final state, a neutral B meson which at time t = 0 is a pure $|B_d\rangle$ oscillates between being a $|B_d\rangle$ and a $|\overline{B_d}\rangle$. This oscillation has been observed,[6] and it is found that

$$\frac{\Delta m}{\Gamma} = \begin{cases} 0.66 \pm 0.09 & \text{ARGUS \& CLEO}^{7} \\ 0.72 \pm 0.04 & \text{LEP}^{6,8} \end{cases}$$

Thus, before a typical B decays, it undergoes a non-negligible fraction of one oscillation.

Suppose, now, that f is a final state into which both a pure $B_d$ and a pure $\overline{B_d}$ can decay. Examples of such a final state are $\rho^+\pi^-$, $D^0K_s$, $\pi^+\pi^-$, and $\Psi K_s$. Let $\Gamma_f(t) \equiv \Gamma(B_d(t) \to f)$ be the time-dependent probability for the time-evolved particle $B_d(t)$, which at t = 0 was a pure $B_d$, to decay into f. From the wave function for $B_d(t)$, Eq. (4.12), $\Gamma_f(t)$ is given by

$$\Gamma_f(t) = \left|\langle f|T|B_d(t)\rangle\right|^2 = e^{-\Gamma t}\left|c\langle f|T|B_d\rangle - ie^{-2i\delta^m_{CKM}} s\langle f|T|\overline{B_d}\rangle\right|^2 \ . \tag{4.16}$$

Let us now assume that the decay amplitudes $\langle f|T|B_d\rangle$ and $\langle f|T|\overline{B_d}\rangle$ are each dominated by a single Feynman diagram. Then

$$\langle f|T|B_d\rangle = Me^{i\delta^f_{CKM}}e^{i\alpha_S} \ , \tag{4.17}$$

where M is the magnitude of the diagram which dominates $\langle f|T|B_d\rangle$, $\delta^f_{CKM}$ is the phase of the product of CKM elements to which this diagram is proportional, and $\alpha_S$ is the strong interaction phase of the diagram. Similarly,

$$\langle f|T|\overline{B_d}\rangle = \overline{M}e^{-i\overline{\delta}^f_{CKM}}e^{i\overline{\alpha}_S} \ , \tag{4.18}$$

where $\overline{M}$, $-\overline{\delta}^{f}_{CKM}$, and $\overline{\alpha}_S$ are respectively the magnitude, CKM phase, and strong phase of the diagram which dominates $\langle f | T | \overline{B_d} \rangle$. From Eqs. (4.16)-(4.18), we then have[9]

$$\Gamma_f(t) = e^{-\Gamma t} \{c^2 M^2 + s^2 \overline{M}^2 - 2cs M\overline{M} \sin(\varphi + \varphi_s)\} \quad , \tag{4.19}$$

where

$$\varphi \equiv 2\delta^m_{CKM} + \delta^f_{CKM} + \overline{\delta}^f_{CKM} \tag{4.20}$$

is the relative CKM phase of the two interfering amplitudes in Eq. (4.16), and

$$\varphi_S = \alpha_S - \overline{\alpha}_S \tag{4.21}$$

is their relative strong phase.

The CP-mirror image of the decay $B_d(t) \to f$ is the process $\overline{B_d}(t) \to \overline{f}$, where $\overline{B_d}(t)$ is the time-evolved particle which at time t = 0 is a pure $\overline{B_d}$. As before, when we go from a process to its CP-mirror image, the CKM phases reverse, but nothing else changes. Thus, from the expression (4.19) for $\Gamma_f(t)$, we may infer that the probability $\overline{\Gamma}_{\overline{f}}(t) \equiv \Gamma(\overline{B_d}(t) \to \overline{f})$ is given by[9]

$$\overline{\Gamma}_{\overline{f}}(t) = e^{-\Gamma t} \{c^2 M^2 + s^2 \overline{M}^2 - 2cs M\overline{M} \sin(-\varphi + \varphi_s)\} \quad . \tag{4.22}$$

Now, since $B_d(t) \to f$ and $\overline{B_d}(t) \to \overline{f}$ are CP-conjugate reactions, CP invariance would require that $\Gamma_f(t) = \overline{\Gamma}_{\overline{f}}(t)$. Comparing Eqs. (4.19) and (4.22), we see that when $\varphi \neq 0$, this requirement is violated. Note that, as always, the CKM phase $\varphi$ produces this CP violation through an interference; in this case the interference between the two terms in Eq. (4.16), or between their analogues in $\overline{B_d}(t) \to \overline{f}$. Physically, the first term in Eq. (4.16) corresponds to a $B_d$ remaining a $B_d$ and decaying directly into f. The second term corresponds to a $B_d$ evolving, through mixing, into a $\overline{B_d}$, which then decays into f.

Recalling that $\Gamma$ and $\Delta m$ are already known, it is trivial to see from Eqs. (4.19) and (4.22) that measurements of the functions $\Gamma_f(t)$ and $\overline{\Gamma}_{\overline{f}}(t)$ will determine M, $\overline{M}$, $s_+ \equiv \sin(\varphi + \varphi_s)$ and $s_- \equiv \sin(-\varphi + \varphi_s)$. Once $s_+$ and $s_-$ are known, one can find $\sin^2\varphi$, up to a two-fold ambiguity, by using

$$\sin^2\varphi = \frac{1}{2}\left[1 - s_+ s_- \pm \sqrt{(1-s_+^2)(1-s_-^2)}\right] \quad . \tag{4.23}$$

Note that, apart from the discrete ambiguity, this expression gives a theoretically *clean* value for $\sin^2\varphi$. This value does not depend on any unknown or difficult-to-calculate parameters. This value can be compared directly to the prediction from the CKM matrix to test cleanly whether phases in this matrix are indeed the source of CP violation.

As we have seen, the CKM phase $\varphi$ which is probed in a given decay, $B_d(t) \to f$, is the relative CKM phase of the two interfering terms in Eq. (4.16). That is, recalling Eq. (4.6),

$$\varphi = \text{CKM Phase}\left[\frac{A(B_d \to f)}{A(B_d \to \overline{B_d})\, A(\overline{B_d} \to f)}\right] , \tag{4.24}$$

where "A" denotes an amplitude. As an example, in $B_d(t) \to \rho^+\pi^-$, we expect $A(B_d \to \rho^+\pi^-)$ to be dominated by the diagram in Fig. 3. Similarly, we expect $\overline{B_d} \to \rho^+\pi^-$ to be dominated by the diagram in Fig. 4.

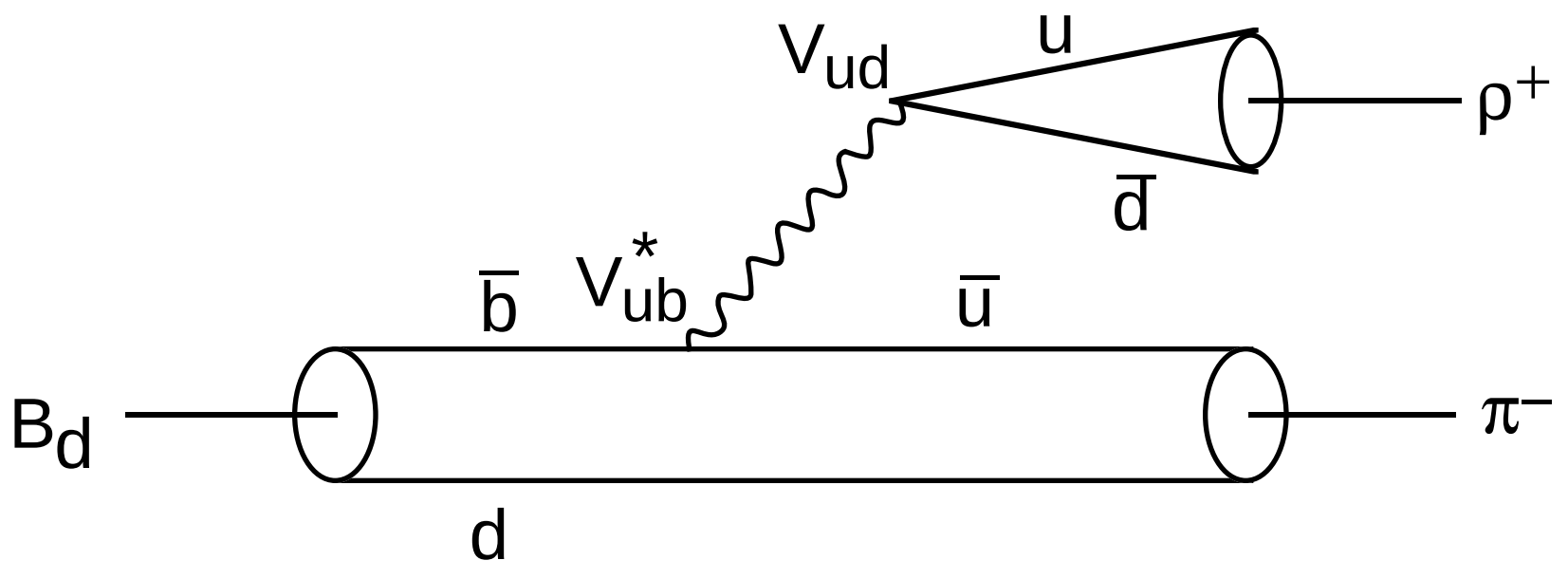

Figure 3. The diagram which dominates $B_d \to \rho^+\pi^-$.

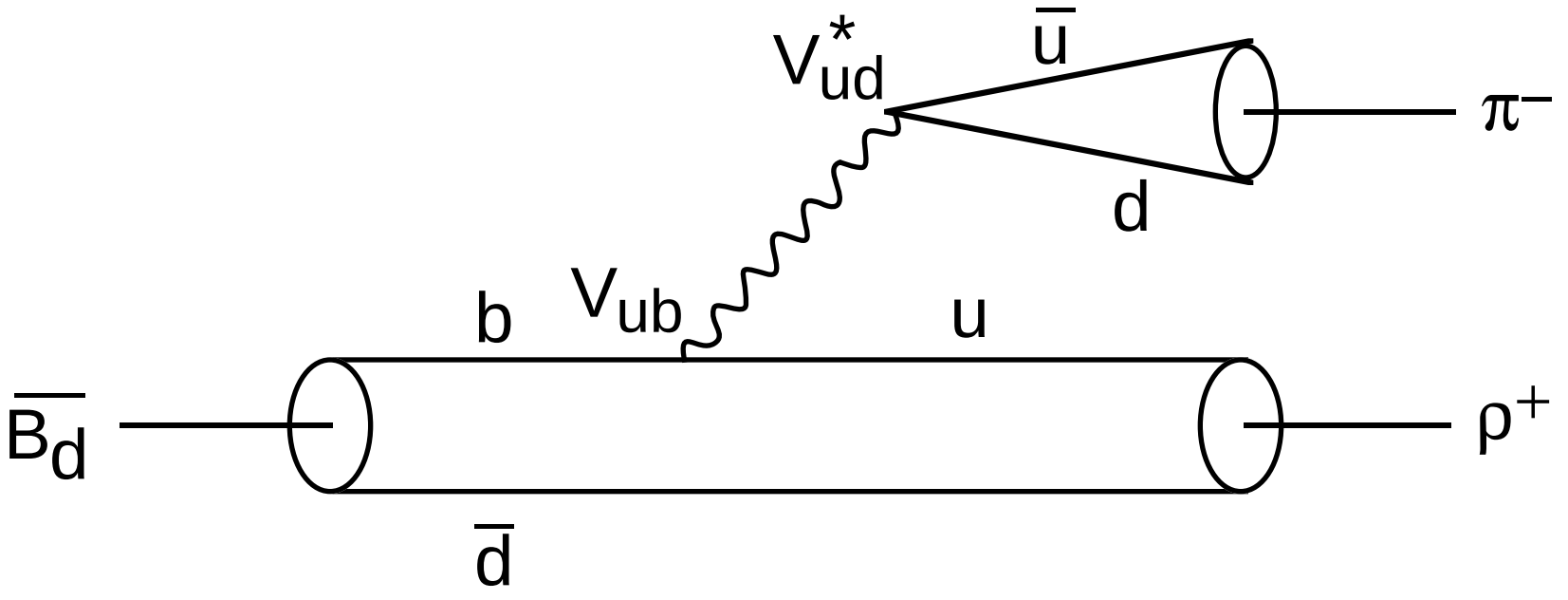

Figure 4. The diagram which dominates $\overline{B_d} \to \rho^+\pi^-$.

The mixing amplitude $A(B_d \to \overline{B_d})$ is dominated by the diagram in Fig. 2. Thus, in $B_d(t) \to \rho^+\pi^-$,

$$\begin{aligned}\varphi &= \arg\left[\frac{V_{ud}V_{ub}^*}{\left(V_{td}V_{tb}^*\right)^2\ V_{ub}V_{ud}^*}\right] \\ &= 2\arg\left[V_{ud}V_{ub}^*V_{tb}V_{td}^*\right]\end{aligned} \quad . \tag{4.25}$$

In a similar way, one may easily find what CKM phase $\varphi$ is probed by any particular decay. Note that since each of the amplitudes in Eq. (4.24) is always proportional to some product of CKM elements (assuming each amplitude is dominated by one diagram), $\varphi$ is always the phase of some product and quotient, or equivalently of some product, of CKM elements.

The neutral B decay rates, and the extraction of a CKM phase from them, become particularly simple when the final state f is a CP eigenstate. Examples of such a final state are $\pi^+\pi^-$ and (neglecting CP violation in the kaon system) $\Psi K_S$. When f is a CP eigenstate, we have $|\bar{f}\rangle \equiv CP|f\rangle = \eta_f|f\rangle$, where $\eta_f$ is the CP parity of $|f\rangle$. Then $\langle f|T|\overline{B_d}\rangle = \eta_f\langle\bar{f}|T|\overline{B_d}\rangle$. Now, $\overline{B_d}$ is the CP conjugate of $B_d$, and $\bar{f}$ is the CP conjugate of f, so $\langle\bar{f}|T|\overline{B_d}\rangle$ is the CP conjugate of $\langle f|T|B_d\rangle$. As before, CP-conjugate amplitudes have opposite CKM phase but are otherwise identical. Thus, from Eq. (4.17), when f is a CP eigenstate,

$$\left\langle f|T|\overline{B_d}\right\rangle = \eta_f M e^{-i\delta^f_{CKM}} e^{i\alpha_S} \quad . \tag{4.26}$$

Using this relation and Eq. (4.17) in Eq. (4.16), we find that

$$\Gamma_f(t) = M^2 e^{-\Gamma t}\{1 - \eta_f \sin\varphi \sin(\Delta m\, t)\} \quad , \tag{4.27}$$

where $\varphi$, the relative CKM phase of the two interfering terms, is now given by

$$\varphi = 2\left(\delta^m_{CKM} + \delta^f_{CKM}\right) \quad . \tag{4.28}$$

For the CP-mirror-image decay, $\overline{B_d}(t) \to f$, the decay rate $\bar{\Gamma}_f(t)$ must be the same as $\Gamma_f(t)$ except for a reversal of the CKM phase. That is,

$$\bar{\Gamma}_f(t) = M^2 e^{-\Gamma t}\{1 + \eta_f \sin\varphi \sin(\Delta m\, t)\} \quad . \tag{4.29}$$

Now, $\Delta m$ is known, as is the CP parity $\eta_f$ of any particular final state f of interest. Thus, the CP-violating asymmetry between $\bar{\Gamma}_f(t)$ and $\Gamma_f(t)$,

$$\frac{\bar{\Gamma}_f(t) - \Gamma_f(t)}{\text{''} \quad + \quad \text{''}} = \eta_f \sin\varphi \sin\left(\Delta m\, t\right) \quad , \tag{4.30}$$

cleanly determines the CKM phase quantity sinφ.[10]

It should be emphasized that the ability to cleanly extract CKM phase information from decay rates does depend on the assumption that $\langle f|T|B_d\rangle$ and $\langle f|T|\overline{B_d}\rangle$ are each dominated by one Feynman diagram. When $\langle f|T|B_d\rangle$ or $\langle f|T|\overline{B_d}\rangle$ involves several competing diagrams with different CKM phases, the rate for $B_d(t) \to f$ involves several interferences, rather than just one, and no longer cleanly determines any one relative CKM phase of two amplitudes. Fortunately, in at least some of the decay modes of greatest interest, there are strong reasons for believing that one diagram does dominate.[11]

## 4.2. Future Experiments

In Section 3, it was argued that CP-violating effects in B decay can be large. We now see, for example, from Eq. (4.30) for the asymmetry in decay to a CP eigenstate, that these effects can indeed be large. If the CKM phase quantity sinφ in the asymmetry (4.30) is $O(1)$, then obviously the asymmetry itself is $O(1)$. However, it will take a large sample of B mesons to observe even a large CP-violating asymmetry. The reason is that each of the asymmetries on which the experimental search will focus occurs in the decay to some specific final state, or CP-conjugate pair of final states, and the branching ratio for B decay to any of the final states of interest is rather small. Thus, a lot of B mesons will be required before a CP-asymmetry in some particular decay mode can be seen.

As an example, consider the CP eigenstate final state $f = \Psi K_s$. If the decay rate $\Gamma[B_d(t) \to \Psi K_s]$ is measured by observing N events, the measurement has a statistical error of order $\sqrt{N}$. Similarly for $\Gamma[\overline{B_d}(t) \to \Psi K_s]$. Thus, if the asymmetry

$$\frac{\Gamma\left[\overline{B_d}(t) \to \Psi K_s\right] - \Gamma\left[B_d(t) \to \Psi K_s\right]}{\text{''} \quad + \quad \text{''}} \tag{4.31}$$

is, for example, of order 0.1, we must have $\sqrt{N} \ll (0.1)N$ in order to measure it with any accuracy. Hence, we require $N \gtrsim 10^3$. Now, typically a $\Psi$ is detected via its decay to $\mu^+\mu^-$ or $e^+e^-$. Since only 12% of $\Psi$ particles decay in this way, we need $\sim 10^4$ $B_d(t) \to \Psi K_s$ events in order to detect $10^3$ of them. Furthermore, $BR(B_d(t) \to \Psi K_s) \cong 4 x 10^{-4}$.[2] Thus, to detect $10^3$ $B_d \to \Psi K_s$ decays, we need $\sim 10^8$ $B_d$ mesons. For other typical decay modes of interest, the number of $B_d$ mesons required is similar. However, the total number of $B_d$ mesons recorded to date at CESR, for example, is only $\sim 5x10^6$.[12] To produce and study enough B mesons to measure CP-violating asymmetries in the B system, future experiments are being planned for hadron facilities, and special high-luminosity $e^+e^-$ colliders ("B factories") are being built at SLAC and KEK. The experiments to be done at the hadron facilities and the B factories will complement each other nicely.

To experimentally compare the rate for $B_d(t) \to f$ with that for $\overline{B_d}(t) \to \bar{f}$ (or, when f is a CP eigenstate, that for $\overline{B_d}(t) \to f$), we must, of course, be able to distinguish a $B_d(t)$ from a $\overline{B_d}(t)$. That is, we must be able to tag the B as having been a pure $B_d$, or a pure $\overline{B_d}$, at some specific time $t = 0$. Several methods for doing this are being considered. Let us briefly review them.

At the B factories, B mesons will be produced in pairs via the reaction

$$e^+e^- \to \Upsilon(4s) \to B_d\overline{B_d} \ . \tag{4.32}$$

Since the $\Upsilon(4s)$ [the upsilon(4s)] has intrinsic spin $S = 1$, and B mesons are spinless, the B pair created in this reaction will be in a p wave. Now, after it is produced, each B meson in the pair will evolve, thanks to mixing, into a coherent mixture of pure $B_d$ and pure $\overline{B_d}$. However, at no time can one have two identical bosons in an antisymmetric state such as a p wave. Thus, if at some time which we shall call $t = 0$, one of the B mesons in the pair decays in a fashion which reveals that at the instant of decay it is, say, a $\overline{B_d}$, then, at the same instant, the other B meson in the pair must be a $B_d$. That is, the decay of the one B at $t = 0$ tags the remaining B as a $B_d(t)$. This type of tagging is an interesting modern application of the quantum mechanical correlation first discussed by Einstein, Podolsky, and Rosen (the EPR effect).

What kind of neutral B decay will reveal that at the instant of decay the parent was a $\overline{B_d}$? An example of such a decay is semileptonic decay, the diagrams for which are shown in Fig. 5. From these diagrams, we see that a positively-charged lepton $\ell^+$ can come only from a $B_d$, and a negatively charged one $\ell^-$ only from a $\overline{B_d}$. Thus, the charge of the lepton tells us whether, at the instant of decay, the parent was a $B_d$ or a $\overline{B_d}$.

A typical B factory experiment might study the decay chain

$$e^+ + e^- \to \Upsilon(4s) \to \underset{\hookrightarrow\ \ell^{(\mp)} + \ldots}{B} \quad + \quad \underset{\hookrightarrow\ f_{CP}}{B} \ , \tag{4.33}$$

where $\ell^{(\mp)} + \ldots$ is a semileptonic final state and $f_{CP}$ is a CP eigenstate. Let us consider this chain in the $\Upsilon(4s)$ rest frame. In this frame the B mesons are quite nonrelativistic, so we may, for the moment, neglect their motion, and take B-rest-frame proper times and $\Upsilon(4s)$-frame times to be indistinguishable. Changing the notation, let us now call the time of the decay Υ(4s) → B + B, t = 0; the time of the decay B → $l^{\overset{-}{(+)}}$ + X, $t_l$; and the time of the decay B → $f_{CP}$, $t_{CP}$. The probability that one B will decay semileptonically at time $t_l$ is proportional to exp[$-\Gamma t_l$]. The probability that the other B will live at least until time $t_l$ is proportional to a second factor of exp[$-\Gamma t_l$]. If the B undergoing the semileptonic decay yields an $l^-$($l^+$), then at time $t_l$ the other B must be a pure $B_d$ ($\overline{B_d}$). Thus, the probability that this B will decay to $f_{CP}$ at time $t_{CP}$ is given by Eq. (4.27) [Eq. (4.29)] with f taken to be $f_{CP}$. Most importantly, in applying Eq. (4.27) or (4.29), we must take the time variable, which as we recall represents the time of the decay to the CP eigenstate relative to the time when the parent was known to be a pure $B_d$ or $\overline{B_d}$ , to be $t_{CP} - t_l$. Combining all factors, we have for the joint probability of the two B decays in (4.33)

$$\text{Probability}\left[\text{One } B \to l^{\overset{-}{(+)}} + X \text{ at time } t_l;\ \text{Other } B \to f_{CP} \text{ at time } t_{CP}\right]$$

$$\propto e^{-\Gamma t_l} e^{-\Gamma t_l} e^{-\Gamma (t_{CP} - t_l)} \left\{ 1 \overset{-}{(+)}\, \eta_{f_{CP}} \sin\varphi \sin\left[\Delta m \left(t_{CP} - t_l\right)\right] \right\} \quad . \quad (4.34)$$

$$= e^{-\Gamma (t_{CP} + t_l)} \left\{ 1 \overset{-}{(+)}\, \eta_{f_{CP}} \sin\varphi \sin\left[\Delta m \left(t_{CP} - t_l\right)\right] \right\}$$

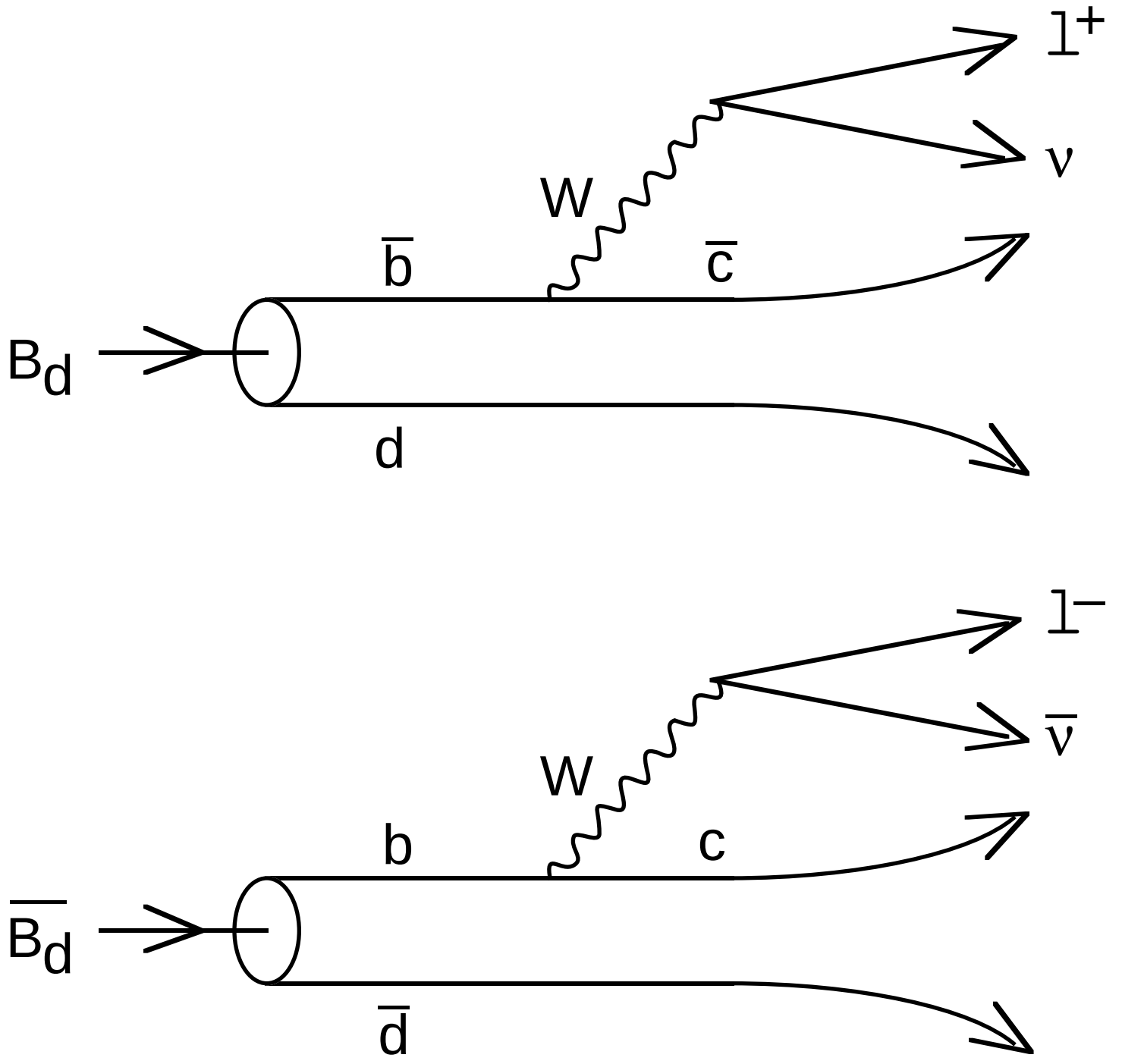

Figure 5. The diagrams for semileptonic neutral B decay.
The symbol l denotes a charged lepton.

Although it is not obvious from what has been said, this result is true even if $t_{CP}$ is earlier than $t_l$.

To take the (so far neglected) motion of the B mesons in the Y(4s) rest frame and all the requirements of relativity into account, we may replace the treatment above by one in which we do not speak of the semileptonic decay of one B as determining the $B_d$ or $\overline{B_d}$ nature of the other B. Rather, we simply calculate directly the amplitude for the entire decay chain (4.33).[13] This amplitude approach also has the advantage of avoiding a puzzling question raised by the treatment based on the EPR effect: How does the second B know the charge of the lepton produced in the decay of the first B, and how does it know when that decay occurred? For the joint probability of the two B decays in (4.33), the amplitude approach yields precisely the same result, Eq. (4.34), as the EPR approach, provided that the times in that result are taken to be *proper* times in the B rest frames, rather than times in the Y(4s) rest frame. The time $t_l$ must be taken to be the proper time elapsed in the frame of the semileptonically decaying B between its birth and decay, and similarly for $t_{CP}$.

Suppose one does an experiment in which there is not enough resolution to measure the decay times $t_l$ and $t_{CP}$, so one simply measures the time integral over the joint decay probability (4.34). The contribution to this time integral of the term in (4.34) proportional to $\sin\varphi$, the quantity one would like to determine, vanishes. This is because

$$\int_0^\infty dt_l \int_0^\infty dt_{CP}\, e^{-\Gamma(t_{CP}+t_l)} \sin[\Delta m(t_{CP}-t_l)] = 0 \qquad (4.35)$$

by the antisymmetry of the integrand under $t_l \leftrightarrow t_{CP}$. Thus, to determine $\sin\varphi$ with neutral B mesons at a B factory, one *must* be able to measure the B decay times, at least to some extent. To measure the decay time of a B, one would determine the pathlength it covers before decay and its energy. Now, in every $e^+e^-$ collider built so far, the $e^+$ and $e^-$ beams have equal and opposite momenta, so that in the reaction $e^+e^- \rightarrow Y(4s) \rightarrow B\overline{B}$, the Y(4s) is at rest in the laboratory frame. Thus, at these colliders, one would be trying to determine the B pathlength in the Y(4s) rest frame. However, as already mentioned, in this frame the B mesons are quite nonrelativistic. In fact, they are so slow ($\beta \cong 0.06$) that, before decaying in $1.6\times10^{-12}$ sec,[8] a typical B covers only ~30μm. Pathlengths this short cannot be measured. To make the B pathlengths long enough to be measurable, the SLAC and KEK B factories will be *asymmetric* colliders. That is, in each of them the positron beam will have a different energy from the electron beam. As a result, the Y(4s) formed in the $e^+e^-$ collision will be moving in the laboratory, and will transmit its motion to its daughter B mesons. The asymmetry between the beam energies will be sufficient to lead to B mesons which typically will travel ~200μm before decaying. Such a distance is large enough to be measured.

Another method for tagging, which may prove useful at hadron facilities, is based on the expectation that some fraction of the neutral B mesons made at those facilities will be created via the production and decay of a B**. By B** we mean an excited B meson heavy enough to decay to B + π. Such mesons are expected as p-wave quark-antiquark bound states, and are observed at LEP.[8] Now, as Fig. 6 makes clear, a $B^{**+}$ decays to $B_d\pi^+$, but a $B^{**-}$ to $\overline{B_d}\pi^-$.

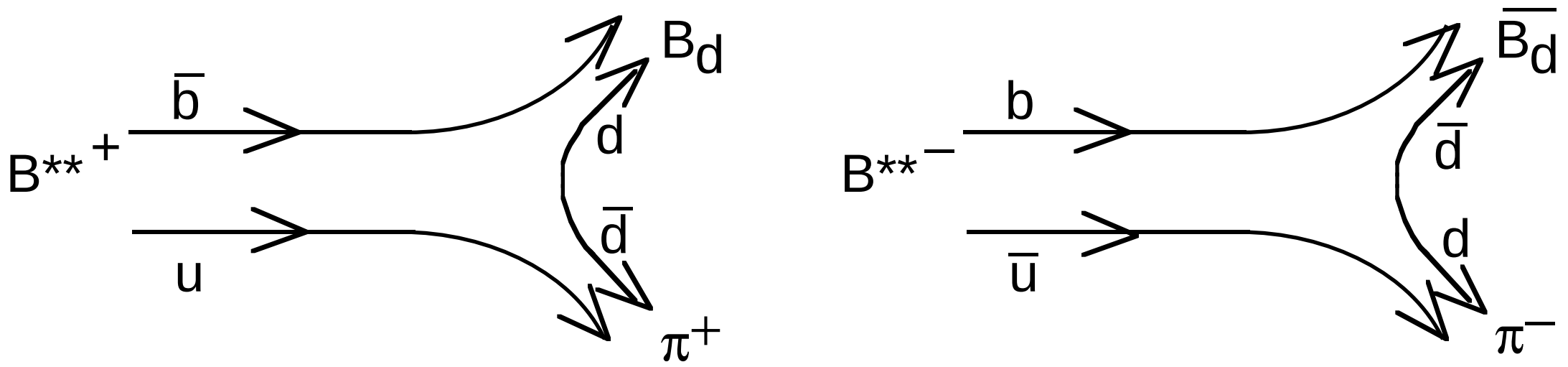

Figure 6. The diagrams for $B^{**+} \to B_d\pi^+$ and $B^{**-} \to \overline{B_d}\,\pi^-$.

Suppose, then, that in some event one finds a neutral B and a charged π which are close to each other in phase space and whose momenta are such that the invariant mass of the Bπ system is the known mass of a B**. Then, neglecting background and assuming that the Bπ system came from a B**, if the charge of the π is positive (negative), we can conclude that, at the moment of its production in $B^{**} \to B\pi$, the neutral B was a pure $B_d$ ($\overline{B_d}$ ).[14] Results from LEP[8] suggest that the fraction of B mesons made via a B** may be appreciable at hadron facilities, so this method of tagging may be quite helpful.

### 4.3. What There is to Measure

As we have seen, the CKM phase φ which is probed by CP violation in any B decay is the phase of some product of CKM elements. What, then, are the independent phases of all possible products of CKM elements? That is, what is there to measure?

The answer to this question grows out of the fact that, in the SM, the CKM matrix must be unitary. The requirement of unitarity demands, among other things, that any pair of columns of the CKM matrix be orthogonal, and similarly for any pair of rows. Thus, we have the six orthogonality constraints

$$
\begin{array}{llr}
\text{ds} & \underset{\lambda}{V_{ud}V^*_{us}} + \underset{\lambda}{V_{cd}V^*_{cs}} + \underset{\lambda^5}{V_{td}V^*_{ts}} = 0 & \\
\text{sb} & \underset{\lambda^4}{V_{us}V^*_{ub}} + \underset{\lambda^2}{V_{cs}V^*_{cb}} + \underset{\lambda^2}{V_{ts}V^*_{tb}} = 0 & \\
\text{db} & \underset{\lambda^3}{V_{ud}V^*_{ub}} + \underset{\lambda^3}{V_{cd}V^*_{cb}} + \underset{\lambda^3}{V_{td}V^*_{tb}} = 0 & (4.36) \\
\text{uc} & \underset{\lambda}{V_{ud}V^*_{cd}} + \underset{\lambda}{V_{us}V^*_{cs}} + \underset{\lambda^5}{V_{ub}V^*_{cb}} = 0 & \\
\text{ct} & \underset{\lambda^4}{V_{cd}V^*_{td}} + \underset{\lambda^2}{V_{cs}V^*_{ts}} + \underset{\lambda^2}{V_{cb}V^*_{tb}} = 0 & \\
\text{ut} & \underset{\lambda^3}{V_{ud}V^*_{td}} + \underset{\lambda^3}{V_{us}V^*_{ts}} + \underset{\lambda^3}{V_{ub}V^*_{tb}} = 0 &
\end{array}
$$

To the left of each constraint is indicated the pair of columns, or of rows, whose orthogonality is expressed by that constraint. Under each term in each constraint is given the rough empirical size of that term, expressed as a power of the Cabibbo angle $\lambda = 0.22$. Each term in any of the constraints may be pictured as a vector in the complex plane. The constraint then states that its three terms form the sides of a closed triangle, called a "unitarity triangle",[15] in the complex plane. The six unitarity triangles corresponding to the constraints of Eqs. (4.36) are shown, somewhat schematically, in Fig. 7. As we see from Eqs. (4.36), in two of the triangles, the three sides are of comparable size, so that the interior angles can all be large. However, in each of the remaining triangles, one of the sides is much shorter than the other two, and the angle opposite this short side must be small.

Any angle in one of the unitarity triangles is, of course, (apart from an extra $\pi$) just the relative phase of the two adjacent sides. Thus, this angle is the phase of a product of CKM elements. Furthermore, the product concerned will always be one whose phase is convention-independent. For example, the relative phase $\psi$ of the two sides adjacent to the angle $\alpha$ in the db triangle is arg $(V_{td}V^*_{tb}V^*_{ud}V_{ub})$. Now, this phase is invariant under phase redefinition of the t quark, since this redefinition causes equal and opposite phase changes in $V_{td}$ and $V^*_{tb}$. Similarly, $\psi$ is invariant under phase redefinition of the u, d, or b quark. Thus, the angles in the unitarity triangles do not depend on phase conventions.

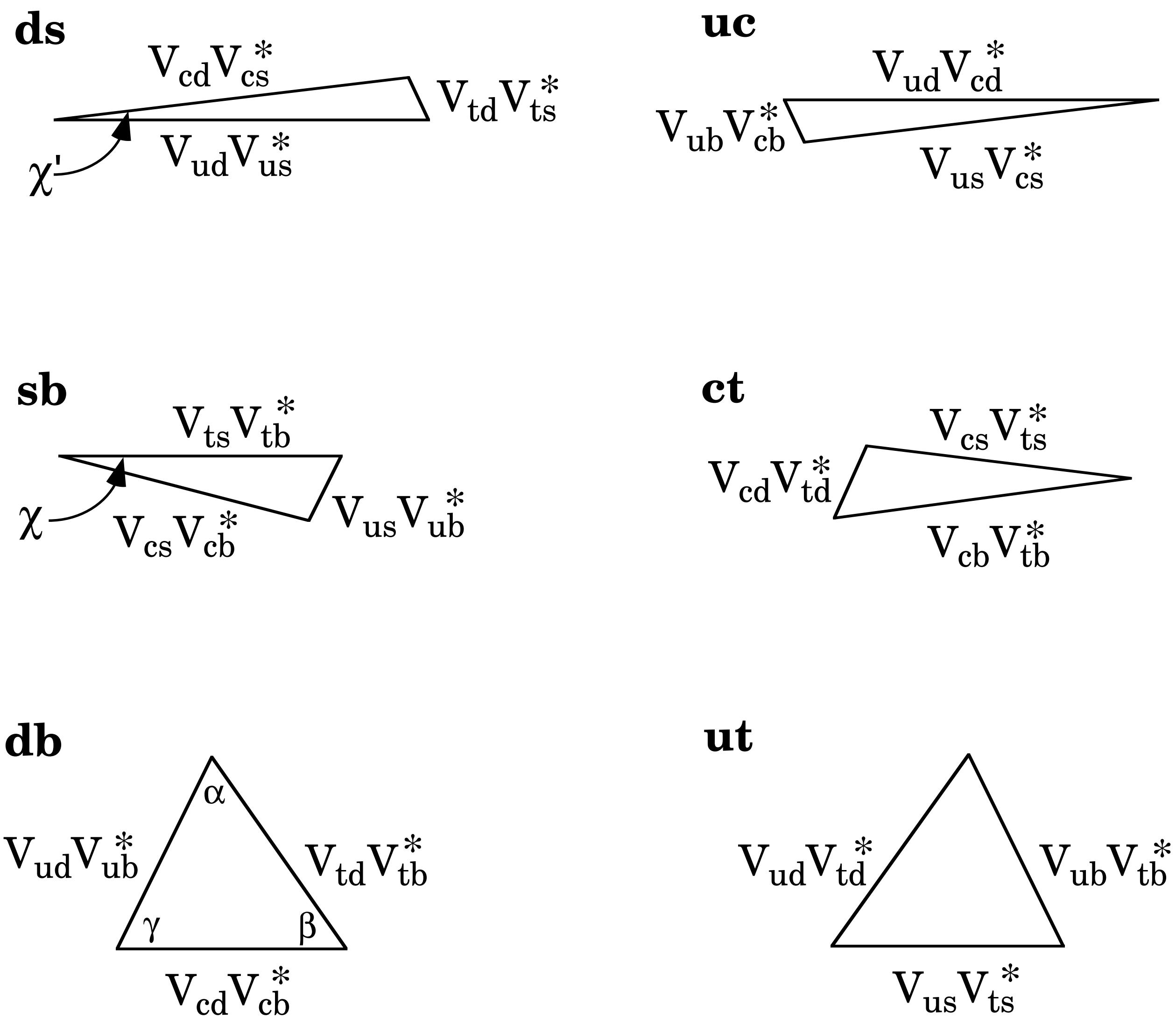

Figure 7. The unitarity triangles. To the left of each triangle is indicated the pair of columns, or of rows, whose orthogonality this triangle expresses. The significance of the angles labeled $\alpha$, $\beta$, $\gamma$, $\chi$, and $\chi'$ is explained in the text.

Now, it can be shown that if $\varphi$ is the phase of any phase-convention-independent product of CKM elements (that is, if $\varphi$ is the CKM phase probed in some experiment on CP violation), then[16]

$$\varphi = n_\alpha \alpha + n_\beta \beta + n_\chi \chi + n_{\chi'} \chi' \ . \tag{4.37}$$

Here, $\alpha$, $\beta$, $\chi$, and $\chi'$ are the four unitarity triangle angles identified in Fig. 7, and $n_\alpha$, $n_\beta$, $n_\chi$, and $n_{\chi'}$ are integers. From Eq. (4.37), we see that, presuming $\alpha$, $\beta$, $\chi$, and $\chi'$ are independent, these four angles may be taken to be the independent phases of all possible (convention-independent) products of CKM elements. The CKM phase φ probed by any CP experiment is a simple linear combination of these four angles. The future experiments on CP violation in the B system may be thought of as, in part, an attempt to determine these four angles.

It can be proved that, once they are known, α, β, χ, and χ′ completely determine the entire CKM matrix.[16] Since, as is well known, it takes four independent parameters to determine this matrix, it follows that α, β, χ, and χ′ must indeed be independent, as we just assumed. Furthermore, since α, β, χ, and χ′ do completely determine the full CKM matrix, CP experiments in the B system are not merely measurements of angles in the unitarity triangles, but, in principle at least, probes of the entire content of the CKM matrix.[17]

From the magnitudes of the terms in the "ds" orthogonality constraint of Eqs. (4.36), we see that the angle χ′ in the ds unitarity triangle is at most of order $\lambda^5/\lambda$, or $2\text{x}10^{-3}$ radians. Thus, in a B decay where the CKM phase φ which is probed is χ′, the CP violation would be very small. As a result, it may not be possible to measure χ′. However, plans are being developed, and facilities being constructed, to measure the three remaining independent angles, α, β, and χ.

Wolfenstein has introduced a very good (~3%) approximation[18] to the CKM matrix V which is based on the empirical observation that, as far as the magnitudes of its elements are concerned, V has approximately the form

$$V \sim \begin{bmatrix} 1 & \lambda & \lambda^3 \\ \lambda & 1 & \lambda^2 \\ \lambda^3 & \lambda^2 & 1 \end{bmatrix} . \tag{4.38}$$

The implications of the magnitudes summarized here for the unitarity constraints (4.36) have already been indicated beneath them. In Wolfenstein's approximation, in effect one neglects the small term in the ds constraint of Eqs. (4.36) relative to the larger terms, and does the same in the sb constraint. The ds and sb unitarity triangles then each collapse to two antiparallel lines of equal length, and the angles χ′ and χ vanish (cf. Fig. 7). Of the four independent unitarity-triangle angles originally present, only the angles α and β, in the db triangle, remain. These angles, and the dependent angle $\gamma = \pi - \alpha - \beta$ in the same triangle, are in any case the angles on which the early CP experiments on the B system will concentrate, since they are the angles which may be large and which, therefore, may produce large CP-violating asymmetries. Consequently, in the literature, attention has been focused on the db triangle.

The program to test the SM of CP violation through experiments on B decays may be summarized as follows:

1. Measure the four independent angles of the unitarity triangles. If the smallest angle, $\chi'$, is beyond reach, at least measure $\alpha$, $\beta$, and $\chi$. Focus first on $\alpha$ and $\beta$, since these angles may both be large.

2. To see whether the SM provides a consistent picture of CP-violating phenomena, or leads to inconsistencies which point to physics beyond the SM, overconstrain the system as much as possible. To do so—

   a. Measure, if possible, CP asymmetries in different decay modes which, *if* the SM of CP violation is correct, all yield the same angle ($\beta$, for example). See whether these asymmetries actually yield the same numerical result.

   b. Measure independently the angles $\alpha$, $\beta$, and $\gamma$ in the db triangle, and see whether these angles actually add up to $\pi$.

   c. Measure the lengths of the sides of the db triangle (via experiments on non-CP-violating effects such as decay rates and neutral B mixing). See whether the interior angles implied by the measured lengths agree with those inferred directly from CP-violating asymmetries.

Table 1. Decay modes and the CKM phase angle $\varphi$ which they probe. In the final state $\Psi K^{*0}$, the $K^{*0}$ is required to decay as shown. Similarly for the final state $\overset{(-)}{D}{}^0 K^+$; $g_{CP}$ is a CP eigenstate, such as $\pi^+\pi^-$ or $K^+K^-$. References are given in the last column.

| Decay Mode | $\varphi$ | Ref. |
|---|---|---|
| $B_d(t) \to \pi^+\pi^-,\ \rho^+\pi^-,\ a_1{}^+\pi^-$ | $2\alpha$ | 11, 9, 19 |
| $B_d(t) \to \Psi K_s,\ \Psi K^{*0}$ ↳ $K_s\,\pi^0$ | $2\beta$ | 10, 20 |
| $B_s(t) \to D_s{}^+K^-$ | $\gamma + 2\chi - \chi'$ | 21 |
| $B^+ \to \overset{(-)}{D}{}^0 K^+$ ↳ $g_{CP}$ | $\gamma - \chi'$ | 5 |
| $B_s(t) \to \Psi\phi$ | $2\chi$ | 22, 23 |

In Table 1 are listed some decay modes which (in combination with their CP conjugates) are potential probes of the independent angles α, β, and χ, and the dependent angle γ. In this table, $B_s(t)$, in analogy with $B_d(t)$, is the time-evolved state which at time t = 0 was a pure $B_s$. The $B^+$ decay listed is one of the exceptional charged B decays from which clean CKM phase information can be extracted.[5] Note that, neglecting χ and χ′ relative to γ, the decays $B_s(t) \to D_s^+K^-$ and $B^+ \to \overset{(-)}{D^0}K^+ \to (g_{CP})K^+$ both yield the latter angle.

## 5. Testing the SM of CP Violation in the K System

The future tests of the SM of CP violation will include experiments on the neutral K system, where CP violation was discovered. Before discussing these experiments, we shall introduce a phase-convention-independent description of CP violation in this system. Such a description has several advantages. First, it clarifies the meaning of the phases which have been experimentally observed. Secondly, it makes possible a useful test for errors in theoretical calculations. Namely, if one computes the theoretical prediction for an experimental observable using nothing but convention-independent variables, then it is easy to check by inspection that the prediction is convention-independent, as it must always be. If it is not convention-independent, then one has made a mistake.

With the convention-independent description of CP violation in hand, we shall discuss past experiments on the kaon analogues of the time-dependent $B_d(t)$ decays we considered in Section 4.1. Finally, we shall turn to future kaon experiments.

### 5.1 Convention-Free Description of CP Violation

The existence of different phase conventions arises from the freedom to redefine any quantum state by multiplying it by a phase factor. To develop a phase-convention-free formalism, we must express every quantity of interest in terms of variables that are manifestly invariant under such phase redefinitions of the states.

When the phases of the states $|K^0\rangle$ and $|\overline{K^0}\rangle$, and in particular their relative phase, are arbitrary, we have

$$\mathrm{CP}\,|K^0\rangle = \omega|\overline{K^0}\rangle\ , \tag{5.1}$$

where ω is a phase factor. Elementary field theory then implies that

$$\mathrm{CP}\,|\overline{K^0}\rangle = \omega^*|K^0\rangle\ . \tag{5.2}$$

Thus, within the neutral K system, in the $K^0$, $\overline{K^0}$ basis, the operator CP is the matrix

$$CP = \begin{bmatrix} 0 & \omega^* \\ \omega & 0 \end{bmatrix} . \tag{5.3}$$

From this matrix, we see that within the neutral K system,

$$(CP)^{-1} = CP = CP^\dagger , \tag{5.4}$$

and

$$(CP)^2 = I , \tag{5.5}$$

where I is the identity matrix. From this last relation, it follows that the neutral kaon CP eigenstates, $|K_{1,2}\rangle$, are given by

$$\left|K_{1(2)}\right\rangle = \frac{e^{i\varphi_{1(2)}}}{\sqrt{2}}\left[\left|K^0\right\rangle \underset{(-)}{+} CP\left|K^0\right\rangle\right] , \tag{5.6}$$

with

$$CP\,|K_{1(2)}\rangle = \underset{(-)}{+}\,|K_{1(2)}\rangle . \tag{5.7}$$

In Eqs. (5.6), the overall phases $\varphi_{1,2}$ are arbitrary. However, when, as in either of Eqs. (5.6), a state is expressed as a coherent superposition of several components, the *relative* phases of the components had better not be arbitrary, because the contributions from these components can interfere, with physical consequences, when the state decays. To make this non-arbitrariness manifest in each of Eqs. (5.6), we have written both components on the right-hand side in terms of the same state, $|K^0\rangle$. It is then obvious that no arbitrary relative phase is involved. (An operator, such as the CP operator in Eq. (5.6), does not introduce arbitrary phases. These come only from states, or from the matrix elements of operators between states.)

Let us now turn to the neutral K mass matrix $\mathcal{M}$ of Eq. (2.4). The diagonal elements of this matrix are convention-free, since the arbitrary phase of the state $|K^0\rangle$ obviously cancels out of $\mathcal{M}_{11} \equiv \langle K^0 | \mathcal{M} | K^0\rangle$ and that of $|\overline{K^0}\rangle$ cancels out of $\mathcal{M}_{22} \equiv \langle\overline{K^0} | \mathcal{M} | \overline{K^0}\rangle$. Thus, the CPT constraint that $\mathcal{M}_{11} = \mathcal{M}_{22} \equiv X$ holds in any convention.

The eigenvalues of $\mathcal{M}$—the complex masses of the mass eigenstates $K_S$ and $K_L$—are

$$\lambda_{S(L)} = X \underset{(-)}{+} \sqrt{\mathcal{M}_{12}\mathcal{M}_{21}} \ . \tag{5.8}$$

We shall prove shortly that, as the notation implies, the eigenvalue $X + \sqrt{\mathcal{M}_{12}\mathcal{M}_{21}}$ ($X - \sqrt{\mathcal{M}_{12}\mathcal{M}_{21}}$) corresponds to the $K_S$ ($K_L$). Being physically observable, these eigenvalues cannot depend on conventions. As we have just seen, X is indeed convention-free, and $\mathcal{M}_{12}\mathcal{M}_{21} = \langle K^0 | \mathcal{M} | \overline{K^0}\rangle\langle\overline{K^0} | \mathcal{M} | K^0\rangle$ clearly does not depend on the phase of any state either.

The eigenstates belonging to the eigenvalues $\lambda_{S(L)}$ are, respectively,

$$\left|K_{S(L)}\right\rangle = \frac{e^{i\varphi_{S(L)}}}{\sqrt{1+|\bar{\rho}|^2}}\left[\left|K^0\right\rangle \underset{(-)}{+} \bar{\rho} CP \left|K^0\right\rangle\right] \ . \tag{5.9}$$

Here, $\varphi_{S(L)}$ are arbitrary phases, and

$$\bar{\rho} \equiv \left[\frac{\left\langle K^0\right|(CP)\mathcal{M}\left|K^0\right\rangle}{\left\langle K^0\right|\mathcal{M}(CP)\left|K^0\right\rangle}\right]^{1/2} \ . \tag{5.10}$$

The arbitrary phase of the state $|K^0\rangle$ obviously cancels out of $\bar{\rho}$, so this quantity is convention-free. Hence, so too is the relative phase of the two terms on the right-hand side of Eqs. (5.9).

In terms of the CP eigenstates, the mass eigenstates $|K_{S(L)}\rangle$ of Eqs. (5.9) are

$$\left|K_{S(L)}\right\rangle = e^{i\varphi_{S(L)}} \frac{1+\bar{\rho}}{\sqrt{2\left(1+|\bar{\rho}|^2\right)}}\left[\left|\tilde{K}_{1(2)}\right\rangle + \bar{\varepsilon}\left|\tilde{K}_{2(1)}\right\rangle\right] \ . \tag{5.11}$$

Here,

$$\left|\tilde{K}_{1(2)}\right\rangle \equiv e^{-i\varphi_{1(2)}}\left|K_{1(2)}\right\rangle \ , \tag{5.12}$$

and

$$\bar{\varepsilon} \equiv \frac{\left\langle K^0\middle|K_1\right\rangle\left\langle K_1\middle|K_L\right\rangle}{\left\langle K^0\middle|K_2\right\rangle\left\langle K_2\middle|K_L\right\rangle} = \frac{1-\bar{\rho}}{1+\bar{\rho}} = \frac{\left\langle K^0\right|\mathcal{M}(CP)\left|K^0\right\rangle^{1/2} - \left\langle K^0\right|(CP)\mathcal{M}\left|K^0\right\rangle^{1/2}}{\left\langle K^0\right|\mathcal{M}(CP)\left|K^0\right\rangle^{1/2} + \left\langle K^0\right|(CP)\mathcal{M}\left|K^0\right\rangle^{1/2}} \ . \tag{5.13}$$

Note that $\bar{\varepsilon}$ is convention-free, and that, from Eqs. (5.12) and (5.6), the same is true of the relative phase of $|\tilde{K}_1\rangle$ and $|\tilde{K}_2\rangle$. Thus, the relative phase of the two terms on the right-hand side of Eqs. (5.11) is independent of conventions.

When the neutral kaon mass matrix $\mathcal{M}$ is CP-invariant, we have $(CP)^{-1}\mathcal{M}(CP) = \mathcal{M}$, so that $\mathcal{M}(CP) = (CP)\mathcal{M}$, and consequently $\bar{\varepsilon}$ vanishes. Thus, $\bar{\varepsilon}$ is a convention-free measure of CP violation in the neutral K mass matrix.

As we noted earlier, CP violation in the neutral K system is small. From the fact that the amplitude for $K_L \to \pi\pi$ is much smaller than that for $K_S \to \pi\pi$ [see Eqs. (2.11) and (2.12)], and the fact that $CP(\pi\pi) = +1$, we know that it is $K_S$ which is close to being a CP-even eigenstate of CP, and $K_L$ which is close to being a CP-odd one. From Eq. (5.13), we see that when CP-noninvariance of $\mathcal{M}$ is small, $\bar{\varepsilon}$ is small. Thus, it is clear from Eq. (5.11) that the mass eigenstates we have labeled "$|K_S\rangle$" and "$|K_L\rangle$" are indeed respectively the $|K_{Short}\rangle$ and $|K_{Long}\rangle$. Hence, the corresponding eigenvalues, "$\lambda_S$" and "$\lambda_L$" of Eq. (5.8), are indeed respectively the complex masses of $K_{Short}$ and $K_{Long}$.

In studying the decays of neutral kaons to a final state f, it will be useful to have the convention-free parameter

$$\overline{\eta_f} \equiv \frac{\langle f|T|K_L\rangle\langle K_L|K^0\rangle}{\langle f|T|K_S\rangle\langle K_S|K^0\rangle} \quad . \tag{5.14}$$

When f is a CP eigenstate with even CP parity, $\overline{\eta_f}$ would vanish in the absence of CP violation, and serves as a convention-free measure of this violation.

In the literature, discussions of CP violation in the kaon system are almost always carried out within specific phase conventions. Almost universally, these discussions adopt the convention that $\varphi_S = \varphi_L = 0$ in Eqs. (5.9) for the states $|K_{S(L)}\rangle$. They also adopt the independent convention that $\varphi_1 = \varphi_2 = 0$ in Eqs. (5.6) for $|K_{(1(2))}\rangle$. Finally, they choose the additional convention that $\omega = +1$ in the CP relation (5.1), as we did in Section 2. Alternatively, they choose $\omega = -1$.

In the literature, neutral kaon decay to the final state f is commonly described in terms of the parameter

$$\eta_f \equiv \frac{\langle f|T|K_L\rangle}{\langle f|T|K_S\rangle} \quad , \tag{5.15}$$

especially when f is a $2\pi$ state. We note that the phase of $\eta_f$ depends on the conventions for the phases of $|K_L\rangle$ and $|K_S\rangle$. Now, from Eqs. (5.9), we see that in the convention where $\varphi_S = \varphi_L$, $\langle K_L | K^0\rangle / \langle K_S | K^0\rangle = 1$. Thus, *in this convention,*

$$\overline{\eta_f} = \eta_f \quad . \tag{5.16}$$

That is, our $\overline{\eta_f}$ is a convention-free analogue of the traditional parameter $\eta_f$, and the two agree in the most commonly used convention for the phases of $|K_L\rangle$ and $|K_S\rangle$.

The violation of CP in the neutral K mass matrix $\mathcal{M}$ is traditionally described in terms of the convention-dependent parameter $\varepsilon$, which may be defined by

$$\varepsilon \equiv \frac{\langle K_1 | K_L\rangle}{\langle K_2 | K_L\rangle} \quad . \tag{5.17}$$

When $\mathcal{M}$ is CP-invariant, $K_L$ has no CP-even (i.e., $K_1$) component, so $\varepsilon$ vanishes. From Eqs. (5.13) and (5.6),

$$\varepsilon = e^{i(\varphi_2 - \varphi_1)}\ \overline{\varepsilon} \quad . \tag{5.18}$$

Thus, $\overline{\varepsilon}$ is a convention-free analogue of $\varepsilon$, and in the popular phase convention where $\varphi_2 = \varphi_1 = 0$, the two agree.[24]

### 5.2. Some Existing Observations of CP Violation in the K System

In Section 2, we already mentioned two CP-violating effects which have been seen in neutral kaon decay. The first of these is the decay of $K_L$, which in the absence of CP violation would have CP = –1, to $\pi\pi$, which has CP = +1. Since $\langle K_L | K^0\rangle / \langle K_S | K^0\rangle$ is just a phase factor [see Eqs. (5.9)], we see from Eqs. (2.11) and (2.12) that the magnitudes of $\overline{\eta_{+-}} \equiv \overline{\eta_{\pi^+\pi^-}}$ and $\overline{\eta_{oo}} \equiv \overline{\eta_{\pi^0\pi^0}}$ are both approximately $2.28 \times 10^{-3}$, and, within errors, are equal. The second CP-violating effect we mentioned is the charge asymmetry $\delta$ of Eq. (2.13).

There is a third observed CP-violating effect, closely related to the decay $K_L \to \pi\pi$, and to the non-exponential decays of $B_d(t)$ mesons to CP eigenstates described by Eq. (4.27). This effect is found in the decay $K^0(t) \to f$ of a time-evolved neutral K, which at time t = 0 was a pure $K^0$, into the final state $f = \pi^+\pi^-$ or $f = \pi^0\pi^0$. Now, the $|K_N\rangle$ (N = S or L) mass eigenstate component of a $K^0$ evolves in time t into $|K_N\rangle\exp(-i\lambda_N t)$. From this fact and Eqs. (5.9) and (5.14), it is trivial to show that the time-dependent probability for the decay $K^0(t) \to f$, $\Gamma(K^0(t) \to f)$, is given by

$$\Gamma\left(K^0(t) \to f\right) \propto e^{-\Gamma_S t} + \left|\overline{\eta}_f\right|^2 e^{-\Gamma_L t} + \\ + 2\left|\overline{\eta}_f\right| e^{-\frac{1}{2}(\Gamma_S+\Gamma_L)t} \cos\left(\Delta m_K t - \overline{\varphi}_f\right) \quad . \tag{5.19}$$

Here, we have written the complex mass $\lambda_N$ of $K_N$ as $m_N - i\Gamma_N/2$, where $m_N$ is the mass of $K_N$ and $\Gamma_N$ is its width. The mass difference $\Delta m_K$ is defined as $m_L - m_S$, and $\overline{\varphi}_f$ is the phase of $\overline{\eta}_f$. Note from Eq. (5.19) that because both the $K_S$ and $K_L$ components of a $K^0(t)$ can decay into $\pi\pi$ (in violation of CP), the rate for $K^0(t) \to \pi\pi$ receives a contribution from the decay of the $K_S$ component, another from that of the $K_L$ component, and a third from an interference term.

A fourth observed CP-violating effect, very similar to the one found in $K^0(t) \to f$, is seen in the decay of neutral kaons produced by a regenerator. The regenerator is a slab of material on which is incident a pure $K_L$ beam—a neutral K beam from which the $K_S$ component has long since decayed away. The regenerator recreates a $K_S$ component in this beam. It is able to do so because a $K_L$ is a coherent superposition of $K^0$ and $\overline{K^0}$, and the amplitudes for the latter two particles to scatter in a material medium differ. Thus, what emerges from the medium will be a different $K^0$–$\overline{K^0}$ superposition from the one which was incident. That is, the emerging kaon beam will contain a $K_S$ component. In particular, if a kaon enters the regenerator as a pure $|K_L\rangle$, it will emerge in the state $|K_r\rangle$ given by

$$|K_r\rangle = |K_L\rangle\langle K_L R|T|K_L R\rangle + |K_S\rangle\langle K_S R|T|K_L R\rangle \quad . \tag{5.20}$$

Here, R stands for the regenerator, so that $\langle K_{L(S)} R|T|K_L R\rangle$ is the amplitude for the regenerator to emit a $K_L$ ($K_S$) when a $K_L$ is incident. Now, after a time t in the rest frame of the kaon $|K_r\rangle$, its $|K_N\rangle$ (N = L or S) mass eigenstate component will have evolved into $\exp(-i\lambda_N t)|K_N\rangle$. Thus, the $|K_r\rangle$ will have evolved into the state $|K_r(t)\rangle$ given by

$$|K_r(t)\rangle = e^{-i\lambda_L t}|K_L\rangle\langle K_L R|T|K_L R\rangle + e^{-i\lambda_S t}|K_S\rangle\langle K_S R|T|K_L R\rangle \quad . \tag{5.21}$$

Omitting an irrelevant overall constant, the amplitude for this time-evolved kaon to decay to the final state f, $\langle f|T|K_r(t)\rangle$, is just

$$\langle f|T|K_r(t)\rangle \propto \overline{\eta}_f\, e^{-i\lambda_L t} + \overline{r} e^{-i\lambda_S t} \quad . \tag{5.22}$$

Here, $\overline{r}$ is the convention-free $K_S$ regeneration amplitude defined by

$$\bar{r} \equiv \frac{\langle K_S R|T|K_L R\rangle}{\langle K_L R|T|K_L R\rangle} \frac{\langle K_L|K^0\rangle}{\langle K_S|K^0\rangle} \quad . \tag{5.23}$$

From Eq. (5.22), the probability $\Gamma(K_r(t) \to f)$ for a neutral kaon to decay to a final state f at a proper time t after emerging from a regenerator is given by

$$\Gamma(K_r(t) \to f) \propto |\bar{r}|^2 e^{-\Gamma_S t} + |\bar{\eta}_f|^2 e^{-\Gamma_L t} + \\ + 2|\bar{r}|\,|\bar{\eta}_f|\, e^{-(\Gamma_S+\Gamma_L)t/2} \cos(\Delta m_K t + \bar{\varphi}_r - \bar{\varphi}_f) \quad . \tag{5.24}$$

Here, $\bar{\varphi}_r$ is the phase of $\bar{r}$. If f is a $\pi\pi$ state (hence CP-even), only the first term in Eq. (5.24) would be present were it not for CP violation.

Through experimental studies of $K_S - K_L$ interference terms such as those in $\Gamma(K^0(t) \to f)$, Eq. (5.19), and $\Gamma(K_r(t) \to f)$, Eq. (5.24), we have learned that[25]

$$\Delta m_K = (3.4894 \pm 0.0073)\ \mu\text{eV} \quad , \tag{5.25}$$

that[25]

$$\bar{\varphi}_{+-} \equiv \arg(\bar{\eta}_{+-}) = (43.56 \pm 0.56)^\circ \quad , \tag{5.26}$$

and that[2]

$$\bar{\varphi}_{oo} \equiv \arg(\bar{\eta}_{oo}) = (43.5 \pm 1.0)^\circ \quad . \tag{5.27}$$

In the literature, the numbers quoted in Eqs. (5.26) and (5.27) are referred to, respectively, as "the phase of $\eta_{+-}$" and "the phase of $\eta_{oo}$". In the most popular phase convention, in which $\eta_f = \bar{\eta}_f$, these numbers do have this significance. However, they do *not* have this meaning in general, since, as we have noticed, the phase of $\eta_f$, Eq. (5.15), depends on conventions. The convention-free quantities whose phases, in any convention, have the values quoted in Eqs. (5.26) and (5.27) are, respectively, $\bar{\eta}_{+-}$ and $\bar{\eta}_{oo}$.

### 5.3. Indirect and Direct CP Violation

There are two ways in which CP can be violated in neutral K decay. First, it can be violated as a consequence of the CP-noninvariance of the neutral K mass matrix, which causes the mass eigenstates $K_S$ and $K_L$ to deviate slightly from being pure CP eigenstates. When the $K_L$, while dominantly the CP-odd state $K_2$, contains a small admixture of the CP-even state $K_1$, as in Eq. (5.11), it can decay to the CP-even state $\pi^+\pi^-$ through its $K_1$ component. It can do this even if the actual K decay amplitudes conserve CP, so that $\langle\pi^+\pi^-|T|K_2\rangle = 0$.

The violation of CP stemming from the fact that $K_S$ and $K_L$ are not CP eigenstates is called "indirect CP violation".

The other way in which CP can be violated is through the decay amplitudes themselves. Examples of possible CP violations in K decay amplitudes would be a nonvanishing value of the CP-changing decay amplitude $\langle\pi^+\pi^-|T|K_2\rangle$, or a nonvanishing value of the difference $\langle\pi^-l^+\nu|T|K^0\rangle - \langle\pi^+l^-\bar{\nu}|T|\overline{K^0}\rangle$ between the amplitudes for two CP-mirror-image processes.

The violation of CP in decay amplitudes themselves is called "direct CP violation".

Suppose that $f_+$ is a CP-even final state. Suppose further that there is no direct CP violation. Then $\langle f_+|T|\tilde{K}_2\rangle = 0$. Thus, from Eqs. (5.14), (5.11), and (5.9),

$$\overline{\eta_{f_+}} = \frac{\bar{\varepsilon}\langle f_+|T|\tilde{K}_1\rangle}{\langle f_+|T|\tilde{K}_1\rangle} = \bar{\varepsilon} \quad . \tag{5.28}$$

That is, when there is no direct CP violation, the parameters $\overline{\eta_{f_+}}$ for different CP-even final states f+ are all equal. In particular, they are all equal to $\bar{\varepsilon}$. Now, Eq. (5.13) clearly implies that

$$\bar{\varepsilon} = \frac{\langle K^0|[\mathcal{M},CP]|K^0\rangle}{\left\{\left[\langle K^0|\mathcal{M}(CP)|K^0\rangle\right]^{1/2} + \left[\langle K^0|(CP)\mathcal{M}|K^0\rangle\right]^{1/2}\right\}^2} \quad . \tag{5.29}$$

This expression makes it particularly obvious that $\bar{\varepsilon}$ vanishes when $\mathcal{M}$ is CP invariant. Since $\bar{\varepsilon}$ is small, the two terms in the denominator D of Eq. (5.29) are approximately equal. Thus, from Eqs. (5.3) and (5.8),

$$D \cong 4\sqrt{\mathcal{M}_{12}\mathcal{M}_{21}} = 2\,(\lambda_S - \lambda_L) \quad . \tag{5.30}$$

The numerator N of Eq. (5.29), being convention-independent, may be evaluated in the convention where the $\omega$ of the CP relation (5.1) is unity. In this phase convention,

$$N = \mathcal{M}_{12} - \mathcal{M}_{21} \quad . \tag{5.31}$$

Now, it can be shown that in the difference $\mathcal{M}_{12} - \mathcal{M}_{21}$, the dispersive part of the matrix element dominates strongly over the absorptive part.[26] Furthermore, the dispersive part of $\mathcal{M}_{12}$ is real and equal to that of $\mathcal{M}_{21}$, except for CKM elements in the former which are replaced by their complex conjugates in the latter. Thus, N

= $\mathcal{M}_{12} - \mathcal{M}_{21}$ is pure imaginary. If, in particular, arg N = $-\pi/2$, then, from Eq. (5.30),

$$\arg \bar{\varepsilon} = \tan^{-1}\left(\frac{2\Delta m_K}{\Gamma_S - \Gamma_L}\right) = (43.46 \pm 0.08)^\circ \ . \qquad (5.32)$$

In the absence of direct CP violation, this angle (or, for arg N = $+\pi/2$, this angle plus $\pi$) is the predicted phase of $\overline{\eta}_{+-}$ and of $\overline{\eta}_{oo}$. Comparing Eq. (5.32) with Eqs. (5.26) and (5.27), we see that the agreement is superb. We note that in obtaining arg $\bar{\varepsilon}$, we used Eqs.(5.8) for the eigenvalues of $\mathcal{M}$. These equations assume the CPT constraint $\mathcal{M}_{11} = \mathcal{M}_{22} \equiv X$. Thus, the agreement between the phase we calculated for $\bar{\varepsilon}$ and the measured phases of $\overline{\eta}_{+-}$ and $\overline{\eta}_{oo}$ is a test of CPT invariance.

All confirmed CP-violating effects observed to date can be explained in terms of indirect CP violation alone. For example, as we have already remarked, the measured magnitudes of $\overline{\eta}_{+-}$ and $\overline{\eta}_{oo}$ are compatible with equality, as required when there is no direct CP violation. (We shall return to this point.) In addition, the measured value of the charge asymmetry $\delta$, Eq. (2.13), is compatible with the hypothesis that this asymmetry arises purely from indirect CP violation. This hypothesis is expected to be a very good one, since, as illustrated in Fig. 8, in the SM there is only one diagram for the decay $K_L \to \pi^- l^+ \nu$, and, similarly, only one for $K_L \to \pi^+ l^- \bar{\nu}$. The violation of CP arises from phase factors, and these phase factors never produce physical effects unless there is an *interference* between amplitudes proportional to them. When a decay involves only one diagram, hence only one amplitude, there can be no interference. Therefore, the decay amplitude cannot violate CP. That is, there can be no “direct” CP violation.

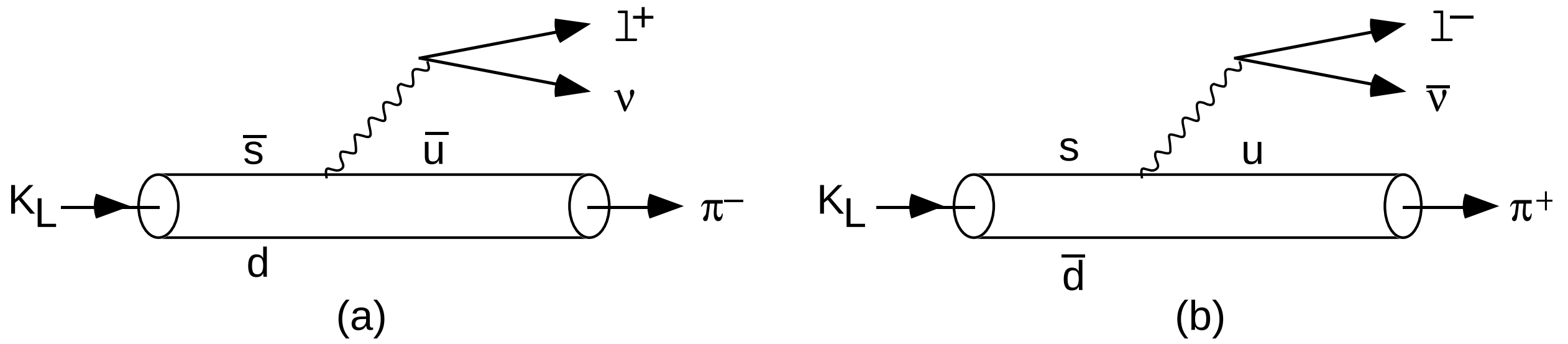

Figure 8. (a) The sole SM diagram for $K_L \to \pi^- l^+ \nu$. (b) The sole SM diagram for $K_L \to \pi^+ l^- \bar{\nu}$. Note that $K_L \to \pi^- l^+ \nu$ proceeds only through the $K^0(\bar{s}d)$ component of the $K_L$, while $K_L \to \pi^+ l^- \bar{\nu}$ proceeds only through the $\overline{K}^0(s\bar{d})$ component.

To see that the value of $\delta$ is compatible with the absence of direct CP violation, we note from Eqs. (5.9) and (5.13) that

$$|K_L\rangle \propto (1+\bar{\varepsilon})|K^0\rangle - (1-\bar{\varepsilon})\omega|\overline{K^0}\rangle \ . \tag{5.33}$$

Recalling (see Fig. 8) that $K_L \to \pi^- l^+ \nu$ and $K_L \to \pi^+ l \bar{\nu}^-$occur only through the $K^0$ and $\overline{K}^0$ components of the $K_L$, respectively, we have

$$\langle \pi^- l^+ \nu | T | K_L\rangle \propto (1+\bar{\varepsilon})\langle \pi^- l^+ \nu | T | K^0\rangle \tag{5.34}$$

and

$$\langle \pi^+ l \bar{\nu} | T | K_L\rangle \propto (1-\bar{\varepsilon})\langle \pi^+ l \bar{\nu} | T | \overline{K^0}\rangle \ . \tag{5.35}$$

If there is no direct CP violation, then $\langle \pi^- l^+ \nu | T | K^0\rangle$ and $\langle \pi^+ l \bar{\nu} | T | \overline{K^0}\rangle$, being decay amplitudes for CP-mirror-image processes, have equal magnitude. Then

$$\begin{aligned} \delta \equiv & \frac{\Gamma(K_L \to \pi^- l^+ \nu) - \Gamma(K_L \to \pi^+ l^- \bar{\nu})}{\text{"} \qquad + \qquad \text{"}} \\ & = \frac{|1+\bar{\varepsilon}|^2 - |1-\bar{\varepsilon}|^2}{\text{"} \quad + \quad \text{"}} \\ & \cong 2\,\Re e\,\bar{\varepsilon} \quad , \end{aligned} \tag{5.36}$$

where we have used the fact that $|\bar{\varepsilon}|^2 \ll 1$. Now, absent direct CP violation, $|\bar{\varepsilon}| = |\overline{\eta}_{+-}| = 2.28 \times 10^{-3}$. Thus, if direct CP violation is also absent from $\delta$, then, from Eq. (5.36), $\delta$ cannot exceed $2\,(2.28 \times 10^{-3}) = 4.56 \times 10^{-3}$. The measured value of $\delta$ quoted in Eq. (2.13) satisfies this constraint easily.

While there is as yet no firm evidence for direct CP violation, a great effort has been made to find such evidence by showing experimentally that in $K \to \pi\pi$, $\overline{\eta}_{oo} \neq \overline{\eta}_{+-}$, in violation of Eq. (5.28). However, so far, this challenging effort has been inconclusive. The reported experimental results are

$$\text{"}\Re e\frac{\varepsilon'}{\varepsilon}\text{"} = \frac{1}{6}\left[1 - \left|\frac{\overline{\eta}_{oo}}{\overline{\eta}_{+-}}\right|^2\right] = \begin{cases} (23 \pm 6.5)\times 10^{-4} & \text{NA31 Experiment}^{27} \\ (7.4 \pm 5.2 \pm 2.9)\times 10^{-4} & \text{E731 Experiment}^{28} \end{cases} \ . \tag{5.37}$$

(In the second of these results, the first error is statistical and the second systematic.) Plainly, more needs to be done to clarify the situation. More sensitive experiments which will try to establish that $\overline{\eta}_{oo} \neq \overline{\eta}_{+-}$ are planned for both Fermilab and CERN. In addition, at the coming $\varphi$ factory DAΦNE, an effort will be made to establish the existence of direct CP violation by following the ingenious suggestion[29] to study the decay chain

$$\varphi \to \underset{\textstyle \hookrightarrow \pi^+\pi^-}{K} \quad + \quad \underset{\textstyle \hookrightarrow \pi^0\pi^0}{K} \quad . \tag{5.38}$$

To see that the probability of this chain depends on whether there is direct CP violation, consider the special case where the two kaons decay simultaneously in the $\varphi$ rest frame. Since the $\varphi$ has S = 1, the primary decay $\varphi \to KK$ leaves the kaons in a p wave. As a result, these two kaons cannot decay simultaneously to the same final state.[30] For, if they did, then just after their decay, we would have two identical spinless bosonic systems (one from each of the kaons) in an overall p wave, in violation of the rule that one cannot have two identical bosons in an antisymmetric state. Thus, if at some time t one of the kaons decays to $\pi^+\pi^-$, then at this time, the other kaon must be that linear combination of $K^0$ and $\overline{K^0}$ which cannot decay to $\pi^+\pi^-$. Now, in the absence of direct CP violation, we have $\langle \pi\pi | T | K_2 \rangle = 0$. Then the linear combination of $K^0$ and $\overline{K^0}$ which cannot decay to $\pi^+\pi^-$ is simply $K_2$. However, (in the absence of direct CP violation) $K_2$ cannot decay to $\pi^0\pi^0$ either. Thus, when there is no direct CP violation, the two kaon decays in the decay sequence (5.38) cannot occur simultaneously.

Of course, the experiment to study the decay chain (5.38) will not restrict itself to events in which the two kaons decay simultaneously. However, by considering this special case, we have seen that the experiment will be sensitive to whether direct CP violation is present or not.

If, as the SM states, CP violation is due to complex phases in the CKM matrix, then direct CP violation is indeed expected to occur, both in K and B decays, apart from exceptions such as $K_L \to \pi^{\mp} l^{\pm} \overset{(-)}{\nu}$. In particular, barring an accident, in $K \to \pi\pi$ the direct CP violation $\langle \pi\pi | T | K_2 \rangle \neq 0$ does indeed occur. Then $\overline{\eta}_{+-} \neq \overline{\eta}_{oo}$, or equivalently, the parameter "$\Re e\,(\varepsilon'/\varepsilon)$" of Eq. (5.37) is nonvanishing. However, calculating the precise SM prediction for $\Re e\,(\varepsilon'/\varepsilon)$ is very challenging. From existing calculations, one predicts only that[31]

$$-2 \times 10^{-4} < \Re e\,(\varepsilon'/\varepsilon) < 13 \times 10^{-4} \ . \tag{5.39}$$

Nevertheless, for $\Re e\,(\varepsilon'/\varepsilon)$ to vanish, or to be much smaller than $10^{-4}$, seems unlikely. Thus, it is very interesting to search, with a sensitivity at the level of $10^{-4}$, for a nonvanishing value of this directly-CP-violating quantity. Establishing a nonvanishing value at this level would not only serve as a test, at least qualitative, of the SM picture of CP violation, but would also discriminate against the models which ascribe CP violation to a so-called "superweak interaction"[32] lying beyond the SM. In general, superweak models of CP violation predict that $\Re e\,(\varepsilon'/\varepsilon) << 10^{-4}$.[32,33]

## 5.4. The Rare Decay $K_L \to \pi^0 \nu\bar{\nu}$

Measurement of the branching ratio for the so far unobserved rare decay $K_L \to \pi^0\nu\bar{\nu}$ would provide a clean test of the SM of CP violation, complementing the tests to come from B decays.

The system $\pi^0\nu\bar{\nu}$ can be in either a CP = +1 or a CP = –1 state. However, neglecting neutrino mass, when this system is produced by SM interactions in $K_L$ decay, it will be in a pure CP = +1 state. But in the absence of CP violation, $CP(K_L) = -1$. Thus, the decay $K_L \to \pi^0\nu\bar{\nu}$ violates CP.

To see why the SM interactions yield a purely CP-even final state in $K_L \to \pi^0\nu\bar{\nu}$, we note that the CP of the final state is given by

$$CP(\pi^0\nu\bar{\nu}) = CP(\pi^0)\, CP(\nu\bar{\nu})\, (-1)^L \ , \tag{5.40}$$

where $CP(\nu\bar{\nu})$ is the CP of the $\nu\bar{\nu}$ pair, and L is the orbital angular momentum of the $\pi^0$ relative to this pair in the $K_L$ rest frame. Since the $K_L$ is spinless, $L = J(\nu\bar{\nu})$, where $J(\nu\bar{\nu})$ is the total angular momentum of the pair. Now, when neutrino mass is neglected, a neutrino produced by SM interactions will be left-handed, and an antineutrino right-handed. Thus, in the effective Hamiltonian $\mathcal{H}_{\mathrm{eff}}$ for $K_L \to \pi^0\nu\bar{\nu}$, in the rest frame of the $\nu\bar{\nu}$ pair, the only operator which can create this pair is $\bar{\nu}_L \vec{\gamma} \nu_L$, where $\nu_L$ is the left-handed projection of the neutrino field. (Other operators bilinear in the neutrino field would create a neutrino pair with the wrong helicities. For example, the scalar operator $\bar{\nu}\nu$ would create a $\nu$ and $\bar{\nu}$ of like, rather than opposite, helicity.) Now, the $\nu\bar{\nu}$ pair created by the action of $\bar{\nu}_L \vec{\gamma} \nu_L$ on the vacuum will have $CP(\nu\bar{\nu}) = +1$, since $\bar{\nu}_L \vec{\gamma} \nu_L$ is even under CP. In addition, this pair will have $J(\nu\bar{\nu}) = 1$, since $\bar{\nu}_L \vec{\gamma} \nu_L$ is a spatial three-vector operator. Thus, since $CP(\pi^0) = -1$, Eq. (5.40) yields $CP(\pi^0\nu\bar{\nu}) = +1$.

In the SM, $K_L \to \pi^0\nu\bar{\nu}$ comes from the diagrams in Fig. 9, plus the related diagrams in which the decay goes through the $\overline{K^0}$, rather than the $K^0$, component of the $K_L$. Notice that all the diagrams in Fig. 9 are proportional to $V_{ts}^* V_{td}$. Thus, their $\overline{K^0}$ analogues, in which every quark has been replaced by its antiquark, are proportional to $V_{ts} V_{td}^*$. Now, from Eqs. (5.9), (5.10), and (5.1),

$$|K_L\rangle \propto \sqrt{\mathcal{M}_{12}}\,|K^0\rangle - \sqrt{\mathcal{M}_{21}}\,|\overline{K^0}\rangle \ . \tag{5.41}$$

At the quark level, the $\overline{K^0} \to K^0$ mixing amplitude $\mathcal{M}_{12}$ arises from diagrams such as, for example, the one in Fig. 10, proportional to $(V_{cs} V_{cd}^*)^2$. Suppose this diagram dominates $\mathcal{M}_{12}$, so that its CP-conjugate, proportional to $(V_{cs}^* V_{cd})^2$, dominates $\mathcal{M}_{21}$. Then, from Eq. (5.41), the diagrams of Fig. 9, and their $\overline{K^0}$ analogues, we have for the $K_L \to \pi^0 \nu\bar{\nu}$ amplitude

$$\langle \pi^0 \nu\bar{\nu} | T | K_L \rangle \propto V_{cs} V_{cd}^* V_{ts}^* V_{td} - V_{cs}^* V_{cd} V_{ts} V_{td}^* \ . \tag{5.42}$$

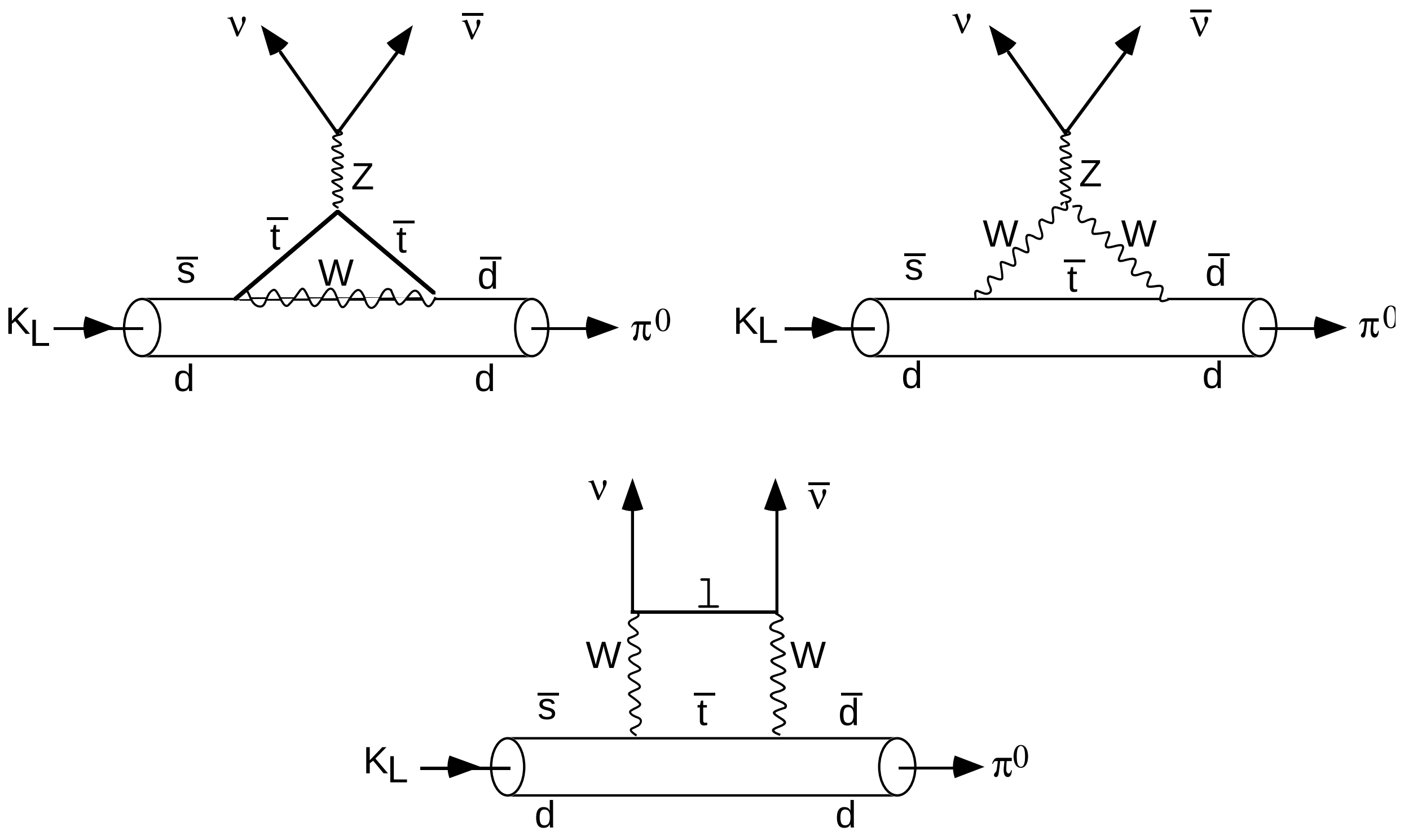

Figure 9. The SM diagrams for $K_L \to \pi^0 \nu\bar{\nu}$ through the $K^0(\bar{s}d)$ component of the $K_L$.

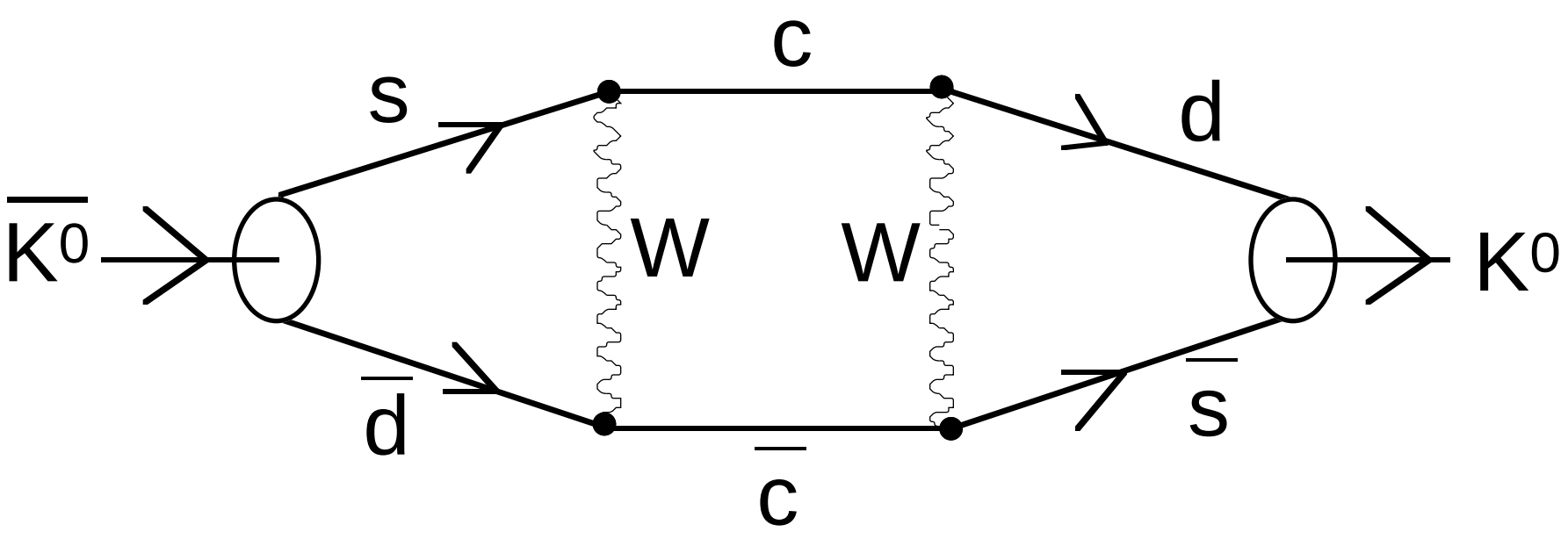

Figure 10. A diagram for $\overline{K^0} \to K^0$ mixing.

That is,

$$\langle \pi^0 \nu\bar{\nu} | T | K_L \rangle \propto \Im m(V_{cs} V_{cd}^* V_{ts}^* V_{td}) \equiv -J \ . \tag{5.43}$$

From Fig. 7, it is easy to see that the phase-convention-independent quantity J is just twice the area of the ds unitarity triangle. What is less obvious is that all six unitarity triangles have the *same* area, so that J, which is known as the Jarlskog invariant, is twice the area of any of them.[34] Since the unitarity triangles can have nonzero area only if the CKM matrix contains CP-violating complex phase factors, J is a convention-free measure of CP violation in the CKM matrix.

In addition to the mixing diagram of Fig. 10, one expects significant contributions from other processes where the quark (antiquark) line involves an intermediate u ($\bar{u}$).[35] These contributions to $\mathcal{M}_{12}$ are proportional to $(V_{us}V_{ud}^*)^2$, rather than $(V_{cs}V_{cd}^*)^2$, and their counterparts in $\mathcal{M}_{21}$ are proportional to $(V_{us}^*V_{ud})^2$. However, in view of the relative sizes of the terms in the ds unitarity constraint of Eqs. (4.36), this constraint implies that, apart from a minus sign, $V_{us}V_{ud}^*$ has the same phase as $V_{cs}V_{cd}^*$ to within a few milliradians. As a result, even when the u-quark contributions to $\mathcal{M}_{12}$ and $\mathcal{M}_{21}$ are included, the CP-violating phase probed by $K_L \to \pi^0\nu\bar{\nu}$ is still that of $V_{cs}V_{cd}^*V_{ts}^*V_{td}$, and we still have $\langle\pi^0\nu\bar{\nu}\,|\,T\,|\,K_L\rangle \propto J$.

The decay $K_L \to \pi^0\nu\bar{\nu}$ is not only CP violating, but is strongly dominated by *direct* CP violation.[36] One way to see this is to estimate $\bar{\eta}_{\pi^0\nu\bar{\nu}}$. To this end, we note that once the heavy W and Z boson degrees of freedom in the diagrams of Fig. 9 are integrated out, the effective Hamiltonian $\mathcal{H}_{eff}$ for $K_L \to \pi^0\nu\bar{\nu}$ is given by[31]

$$\mathcal{H}_{eff} = G\, V_{ts}^*V_{td}\,(\bar{s}\,\gamma_\alpha\, d)\,(\bar{\nu}_L\,\gamma^\alpha\,\nu_L) + \text{h.c.} \quad . \tag{5.44}$$

Here, G is a constant which is fairly well determined once the mass of the top quark is given. For $K_S \to \pi^0\nu\bar{\nu}$, the appropriate effective Hamiltonian represents both the diagrams of Fig. 9 (with $K_L \to K_S$) and similar diagrams with $\bar{t}$ replaced by $\bar{c}$, and so is more complicated than the $\mathcal{H}_{eff}$ of Eq. (5.44). However, we will get the right order of magnitude for the $K_S \to \pi^0\nu\bar{\nu}$ amplitude if we neglect the charm contribution, and take $\mathcal{H}_{eff}$ to be given by Eq. (5.44) for both $K_L \to \pi^0\nu\bar{\nu}$ and $K_S \to \pi^0\nu\bar{\nu}$.[37] Then from the definition (5.14) for $\bar{\eta}_f$, Eq. (5.41) for $|K_L\rangle$ and its analogue for $|K_S\rangle$, and Eq. (5.44) for $\mathcal{H}_{eff}$, we have for $\bar{\eta}_{\pi^0\nu\bar{\nu}}$ the estimate

$$\bar{\eta}_{\pi^0\nu\bar{\nu}} \sim \frac{\left[\sqrt{\mathcal{M}_{12}}V_{ts}^*V_{td}\left\langle\pi^0\middle|\bar{s}\gamma_\alpha d\middle|K^0\right\rangle - \sqrt{\mathcal{M}_{21}}V_{ts}V_{td}^*\left\langle\pi^0\middle|\bar{d}\gamma_\alpha s\middle|\overline{K^0}\right\rangle\right]\left\langle\nu\bar{\nu}\middle|\bar{\nu}_L\gamma^\alpha\nu_L\middle|0\right\rangle}{\text{''} \qquad + \qquad \text{''} \qquad\qquad \text{''}} \quad . \tag{5.45}$$

Now, as we have already noted, the CKM phase of $\sqrt{\mathcal{M}_{12}}$ is that of $V_{cs}V_{cd}^*$. Thus, we may write

$$\sqrt{\mathcal{M}_{12}} = r_{12}V_{cs}V_{cd}^* \quad , \tag{5.46}$$

where $r_{12}$ has no CKM phase. Similarly, we may write

$$\sqrt{\mathcal{M}_{21}} = r_{21} V_{cs}^{*} V_{cd} \ , \tag{5.47}$$

where $r_{21}$ has no CKM phase. Let us now go to a phase convention in which V is real when CP is conserved. In such a convention, we must have

$$r_{12}\left\langle \pi^0 \middle| \bar{s}\gamma_\alpha d \middle| K^0 \right\rangle = r_{21}\left\langle \pi^0 \middle| \bar{d}\gamma_\alpha s \middle| \overline{K^0} \right\rangle \ , \tag{5.48}$$

since $\overline{\eta}_{\pi^0 \nu\bar{\nu}}$ must vanish when CP is conserved. Hence, from Eq. (5.45), in our chosen convention, when CP is *not* conserved, we must have

$$\overline{\eta}_{\pi^0 \nu\bar{\nu}} \sim \frac{i\Im m\left(V_{cs}V_{cd}^{*}V_{ts}^{*}V_{td}\right)}{\Re e\left(V_{cs}V_{cd}^{*}V_{ts}^{*}V_{td}\right)} \ . \tag{5.49}$$

Since both sides of this relation are phase-convention independent, this estimate holds in any convention.

Combined with our knowledge of the CKM matrix,[38] the relation (5.49) yields the estimate $0.1 \lesssim |\overline{\eta}_{\pi^0\nu\bar{\nu}}| \lesssim 1$. Even though this estimate was obtained neglecting the charm-exchange contribution to $K_S \to \pi^0\nu\bar{\nu}$, we may safely conclude that $\overline{\eta}_{\pi^0\nu\bar{\nu}}$ is much larger than the corresponding $K \to \pi\pi$ parameters $\overline{\eta}_{+-}$ and $\overline{\eta}_{00}$, both of which are $\sim 2\text{x}10^{-3}$. However, from Eq. (5.28) we know that when direct CP violation is absent, the parameters $\overline{\eta}_f$ for different CP-even final states are all equal. Thus, we conclude that if the SM description of $K \to \pi^0\nu\bar{\nu}$ is correct, direct CP violation is present in neutral K decays.

From Eqs. (5.14) and (5.11), we have

$$\overline{\eta}_{\pi^0\nu\bar{\nu}} = \frac{A_2 + \bar{\varepsilon}A_1}{A_1 + \bar{\varepsilon}A_2} \ , \tag{5.50}$$

where $A_2$ is the CP-violating decay amplitude $\langle \pi^0\nu\bar{\nu}|T|\tilde{K}_2\rangle$, and $A_1$ is the CP-conserving one $\langle \pi^0\nu\bar{\nu}|T|\tilde{K}_1\rangle$. As we have seen, $|\bar{\varepsilon}| \cong |\overline{\eta}_{+-}| \cong 2\text{x}10^{-3}$, but $|\overline{\eta}_{\pi^0\nu\bar{\nu}}|$ is much larger than this. Thus, from Eq. (5.50), $|A_2| >> |\bar{\varepsilon}A_1|$. That is, in $\langle \pi^0\nu\bar{\nu}|T|K_L\rangle$, which from Eq. (5.11) is proportional to $A_2 + \bar{\varepsilon}A_1$, the directly CP violating term $A_2$ *dominates* over the indirectly CP violating one $\bar{\varepsilon}A_1$.[39]

To calculate $\Gamma(K_L \to \pi^0\nu\bar{\nu})$, one must evaluate the matrix element

$$\langle \pi^0 \nu\bar{\nu} \,|\, \mathcal{H}_{\rm eff} \,|\, K^0 \rangle \;=\; G\, V_{ts}^{*} V_{td}\, \langle \pi^0 \,|\, \bar{s}\, \gamma_\alpha\, d \,|\, K^0 \rangle \, \langle\, \nu\bar{\nu} \,|\, \bar{\nu}_L\, \gamma^\alpha\, \nu_L \,|\, 0 \rangle \qquad (5.51)$$

of the effective Hamiltonian of Eq. (5.44), and the simply-related matrix element $\langle \pi^0 \nu\bar{\nu} | \mathcal{H}_{\rm eff} | \overline{K^0} \rangle$. In Eq. (5.51), the leptonic matrix element is, of course, trivial, and the hadronic matrix element $\langle \pi^0 | \bar{s}\, \gamma_\alpha\, d | K^0 \rangle$ is related by isospin[40] to $\langle \pi^0 | \bar{s}\, \gamma_\alpha\, u | K^+ \rangle$, a matrix element which has been *measured* by determining the rate for $K^+ \to \pi^0 e^+ \nu$. For fixed input parameters (the top quark mass, for example), the theoretical uncertainty in $BR(K_L \to \pi^0 \nu\bar{\nu})$ is only ±1.4%.[31] Thus, measurement of this branching ratio could yield a rather accurate value of the Jarlskog invariant J—information about the CKM matrix which must be consistent with that from other sources.

To be sure, the measurement would not be easy. Given our present knowledge of the relevant CKM elements, $BR(K_L \to \pi^0 \nu\bar{\nu}) = (1\text{-}5)\text{x}10^{-11}$ in the SM. Observing such a rare decay will be a challenge. Adding to the challenge will be the fact that $BR(K_L \to \pi^0\pi^0) \,/\, BR(K_L \to \pi^0 \nu\bar{\nu}) \sim 10^8$, so that one must take steps to ensure that any decay identified as $K_L \to \pi^0 \nu\bar{\nu}$ is not really the much more likely $K_L \to \pi^0\pi^0$, with one of the pions having escaped detection. Nevertheless, it is to be hoped that $BR(K_L \to \pi^0 \nu\bar{\nu})$ will indeed be measured.

## 6. Do Electric Dipole Moments Violate CP?

Outside of the neutral meson systems, CP violation has been sought by trying to show that a spin one-half fermion, such as the neutron or electron, has an electric dipole moment (EDM). It is trivial to prove, as we shall shortly, that such an EDM would violate invariance under time reversal T. *If* one then assumes that the world is invariant under CPT, it follows that an EDM would violate CP. We would like to close the present article by asking whether one can prove directly that an EDM violates CP *without* invoking CPT invariance.

Let us first prove that if a quantum system with definite angular momentum $\vec{s}$ has an EDM, then T invariance is violated. By the rotational properties of the system, the EDM, $\vec{\mu}_{El}$, must point along the vector $\vec{s}$. That is,

$$\vec{\mu}_{El} \;=\; g_{El}\, \vec{s} \;\;, \qquad (6.1)$$

where $g_{El}$ is a constant. Imagine, now, that the system is in a static external electric field $\vec{E}$. The interaction energy $\mathcal{E}$ due to the EDM is then

$$\mathcal{E} \;=\; -\vec{\mu}_{El}\bullet\vec{E} \;=\; -g_{El}\, \vec{s}\bullet\vec{E} \;\; . \qquad (6.2)$$

Now, it is obvious that under time reversal, $\vec{E} \to \vec{E}$, and $\vec{s} \to -\vec{s}$. Thus, $\mathcal{E} \to -\mathcal{E}$. That is, if the interaction energy of the world includes a term stemming from an EDM, then this energy is not invariant under T.

If nature is described by a local, relativistically-invariant quantum field theory (such as the SM and its extensions), then it is invariant under CPT.[41] Thus, it is *relatively* safe to assume that CPT invariance does hold, so that an EDM violates not only T but also CP.

But what if CPT does not hold? Can we prove that an EDM still violates CP? To try to do so, we recall that, apart from a constant, the EDM of a spin one-half fermion f, $\mu_{El}(f)$, is just the $q^2 = 0$ value of the form factor $E(q^2)$ in the "electric dipole term"

$$i\, E(q^2)\, \eta^\mu\, \bar{u}_2\, \sigma_{\mu\nu}\, q^\nu\, \gamma_5\, u_1 \tag{6.3}$$

in the general decomposition of the amplitude for $\gamma + f \to f$, the absorption of a photon by f. Here, q is the momentum carried by the photon, $\eta^\mu$ is the photon polarization, and $u_1$ and $u_2$ are, respectively, the initial and final Dirac spinors for f. To explore the CP properties of the electric dipole coupling, (6.3), let us go to the cross channel, where we have the process $\gamma \to \bar{f} + f$. In this channel, the amplitude (6.3) becomes

$$i\, E(q^2)\, \eta^\mu\, \bar{u}_f\, \sigma_{\mu\nu}\, q^\nu\, \gamma_5\, v_{\bar{f}} \quad , \tag{6.4}$$

where v is the Dirac spinor for an antifermion. We are now in a region where $q^2 \geq (2m_f)^2$, where $m_f$ is the mass of f, while in considering $\gamma + f \to f$ we were in a region where $q^2 \leq 0$. Now, one can show that if $\gamma \to \bar{f} + f$ proceeds through the coupling (6.4), then the $\bar{f}f$ pair will be produced in a state which in the nonrelativistic limit is $^1P_1$. Furthermore, the CP of an $\bar{f}f$ pair in a state with orbital angular momentum L and total spin S is just $(-1)^{S+1}$. Thus, the electric dipole coupling (6.4) leaves the $\bar{f}f$ pair in a state with CP = –1. But a photon has CP = +1. Thus, when $\gamma \to \bar{f} + f$ proceeds through the electric dipole coupling (6.4), CP is not conserved.

We have obtained this result for $\gamma \to \bar{f} + f$ without invoking CPT. Have we thereby proved that an EDM violates CP even when CPT does not hold? Actually, we have not! Recall that the EDM of f is proportional to the value of $E(q^2)$ *at* $q^2 = 0$. To relate $E(q^2{=}0)$ to $E(q^2)$ in the positive $q^2$ region which corresponds to $\gamma \to \bar{f} + f$, we must invoke "crossing". Crossing tells us that the analyticity properties of $E(q^2)$ are such that if this form factor is nonzero in the region $q^2 \leq 0$, which corresponds to $\gamma + f \to f$, then it is also nonzero in the region $q^2 \geq (2m_f)^2$, which corresponds to $\gamma$ $\to \bar{f} + f$. Thus, from our previous argument, if $\mu_{El}(f) \neq 0$, then $\gamma \to \bar{f} + f$ violates CP. However, this author does not know of any way to prove that crossing holds without assuming that nature is described by a local, relativistically-invariant quantum field theory. And, if we assume that nature is described by such a theory, then CPT invariance holds! Thus, we have not succeeded in proving that an EDM violates CP even when CPT does not hold. Indeed, we still do not know whether an EDM necessarily violates CP under such circumstances.

## 7. Conclusion

It is particularly natural to hypothesize that CP violation is an effect of the SM weak interaction, which means that it comes from phases in the quark mixing matrix. During the next two decades, this hypothesis will be cleanly tested through elegant experiments on B and K decays. Regardless of whether these experiments confirm or disprove the hypothesis, their results will be exciting news.

## Acknowledgements

It is a pleasure to thank G. Buchalla, I. Dunietz, B. Holstein, and B. Winstein for helpful conversations, and to thank the Fermi National Accelerator Laboratory for excellent hospitality during the time that these lectures were prepared.